\documentclass[apj]{emulateapj}
\usepackage{graphicx}
\usepackage{multirow}
\usepackage{epsfig}
\newcommand{\sfg}{S$^4$G}
\newcommand{\maggie}{AB\,mag\,arcsec$^{-2}$}
\usepackage{lscape}

\newcommand{\RR}{{\cal R}}
\newcommand{\sub}[1]{\ensuremath{_{\mathrm{#1}}}}

\shorttitle{Disk breaks in \sfg\ galaxies}
\shortauthors{Mu\~{n}oz-Mateos et al.}

\begin{document}

\title{The impact of bars on disk breaks as probed by \sfg\ imaging.}

\author{Juan Carlos Mu\~noz-Mateos\altaffilmark{1},
Kartik Sheth\altaffilmark{1},
Armando Gil de Paz\altaffilmark{2},
Sharon Meidt\altaffilmark{3},
E. Athanassoula\altaffilmark{4},
Albert Bosma\altaffilmark{4},
S\'ebastien Comer\'on\altaffilmark{5},
Debra M. Elmegreen\altaffilmark{6},
Bruce G. Elmegreen\altaffilmark{7},
Santiago Erroz-Ferrer\altaffilmark{8,9},\  
Dimitri A. Gadotti\altaffilmark{10},
Joannah L. Hinz\altaffilmark{11},
Luis C. Ho\altaffilmark{12},
Benne Holwerda\altaffilmark{13},
Thomas H. Jarrett\altaffilmark{14},
Taehyun Kim\altaffilmark{1,10,12,15},
Johan H. Knapen\altaffilmark{8,9},
Jarkko Laine\altaffilmark{5},
Eija Laurikainen\altaffilmark{5,16},  
Barry F. Madore\altaffilmark{12},  
Karin Menendez-Delmestre\altaffilmark{17},
Trisha Mizusawa\altaffilmark{1,18},
Michael Regan\altaffilmark{19},   
Heikki Salo\altaffilmark{5},  
Eva Schinnerer\altaffilmark{3},  
Michael Seibert\altaffilmark{12},
Ramin Skibba\altaffilmark{11},
Dennis Zaritsky\altaffilmark{11}
}

\altaffiltext{1}{National Radio Astronomy Observatory / NAASC, 520 Edgemont Road, Charlottesville, VA 22903; jmunoz@nrao.edu}
\altaffiltext{2}{Departamento de Astrof\'{i}sica, Universidad Complutense de Madrid, Madrid 28040, Spain}
\altaffiltext{3}{Max-Planck-Institut f\"ur Astronomie, K\"onigstuhl 17, 69117 Heidelberg, Germany}
\altaffiltext{4}{Aix Marseille Universite, CNRS, LAM (Laboratoire d'Astrophysique de Marseille) UMR 7326, 13388, Marseille, France}
\altaffiltext{5}{Department of Physical Sciences/Astronomy Division, University of Oulu, FIN-90014, Finland}
\altaffiltext{6}{Vassar College, Dept. of Physics and Astronomy, Poughkeepsie, NY 12604}
\altaffiltext{7}{IBM Research Division, T.J. Watson Research Center, Yorktown Hts., NY 10598}
\altaffiltext{8}{Instituto de Astrof\'{i}sica de Canarias, 38205 La Laguna, Spain}
\altaffiltext{9}{Departamento de Astrof\'{i}sica, Universidad de La Laguna, 38206 La Laguna, Spain}
\altaffiltext{10}{European Southern Observatory, Casilla 19001, Santiago 19, Chile}
\altaffiltext{11}{University of Arizona, 933 N. Cherry Ave, Tucson, AZ  85721}
\altaffiltext{12}{The Observatories, Carnegie Institution of Washington, 813 Santa Barbara Street, Pasadena, CA 91101}
\altaffiltext{13}{European Space Agency, ESTEC, Keplerlaan 1, 2200 AG, Noordwijk, the Netherlands}
\altaffiltext{14}{Astrophysics, Cosmology and Gravity Centre, Department of Astronomy, University of Cape Town, Private Bag X3, Rondebosch 7701, South Africa}
\altaffiltext{15}{Astronomy Program, Department of Physics and Astronomy, Seoul National University, Seoul 151-742, Korea} 
\altaffiltext{16}{Finnish Centre for Astronomy with ESO (FINCA), University of Turku}
\altaffiltext{17}{Observatorio do Valongo, Universidade Federal de Rio de Janeiro, Ladeira Pedro Antöonio, 43, Sa«ude CEP 20080-090, Rio de Janeiro - RJ - Brasil}
\altaffiltext{18}{Florida Institute of Technology, Melbourne, FL 32901}
\altaffiltext{19}{Space Telescope Science Institute, 3700 San Martin Drive, Baltimore, MD 21218}

\begin{abstract}
 
  We have analyzed the radial distribution of old stars in a sample of
  218 nearby face-on disks, using deep 3.6\,$\micron$ images from the
  Spitzer Survey of Stellar Structure in Galaxies (\sfg). In
  particular, we have studied the structural properties of those
  disks with a broken or down-bending profile. We find that, on
  average, disks with a genuine single exponential profile have a
  scale-length and a central surface brightness which are intermediate
  to those of the inner and outer components of a down-bending disk
  with the same total stellar mass. In the particular case of barred
  galaxies, the ratio between the break and the bar radii
  ($R_{\mathrm{br}}/R_{\mathrm{bar}}$) depends strongly on the total
  stellar mass of the galaxy. For galaxies more massive than
  $10^{10}\,M_{\sun}$, the distribution is bimodal, peaking at
  $R_{\mathrm{br}}/R_{\mathrm{bar}} \sim 2$ and $\sim 3.5$. The first
  peak, which is the most populated one, is linked to the Outer
  Lindblad Resonance of the bar, whereas the second one is consistent
  with a dynamical coupling between the bar and the spiral
  pattern. For galaxies below $10^{10}\,M_{\sun}$, breaks are found up
  to $\sim 10 R_{\mathrm{bar}}$, but we show that they could still be
  caused by resonances given the rising nature of rotation curves in
  these low-mass disks. While not ruling out star formation
  thresholds, our results imply that radial stellar migration induced
  by non-axysymmetric features can be responsible not only for those
  breaks at $\sim 2 R_{\mathrm{bar}}$, but also for many of those
  found at larger radii.

\end{abstract}


\section{Introduction}
\setcounter{footnote}{0}
The initial conditions under which galaxies form and the physical
mechanisms that govern their subsequent evolution are encoded in the
stellar structure of galaxy disks. Hence the radial profile of the
stellar disk is a powerful tool for probing galaxy evolution over
cosmic time.

Light profiles of galactic disks have generally been described by an
exponential law (\citealt{Freeman:1970}) with a truncation or a
break\footnote{The terms `break' and 'truncation' are often used
  somewhat interchangeably in the literature. Here we prefer to use
  `break' when talking about changes in slope in the main disk of
  galaxies (the subject of this paper) and leave `truncation' for the
  features seen much further out in edge-on disks (see
  \citealt{Martin-Navarro:2012}).} at the outer edge of the disk
(\citealt{van-der-Kruit:1979}). Subsequent deeper observations showed
that the sharp cutoffs found by van der Kruit are better described by
a double exponential profile (\citealt{Pohlen:2002}), where the slope
of the outer exponential is steeper than that of the inner one; these
are known as down-bending profiles. In contrast to these, some disks
have been observed with an outer disk which has a shallower light
profile than the inner exponential (\citealt{Erwin:2005,
  Pohlen:2006}); these are referred to as up-bending profiles.

The latest data show that systems with a single exponential profile
are the exception rather than the rule in the local Universe
\citep{Erwin:2005, Pohlen:2006}.  By analyzing optical light profiles
of $\sim$90 nearby late-type spirals (Sb-Sdm), \cite{Pohlen:2006}
showed that only 10\% exhibited single exponential profiles (Type~I).
Roughly 60\% were found to have down-bending profiles (Type~II) and
the remaining 30\% showed an up-bending profile
(Type~III). \cite{Erwin:2008} (E08 hereafter) performed a similar
study on a sample of 66 nearby barred, early-type disks (S0-Sb), and
concluded that 27\%, 42\% and 24\% of their galaxies exhibited Type~I,
II and III profiles, respectively; the remaining 6\%-7\% showed a
combination of Type~II and III profiles. Double-exponential profiles
are also common in very late-type systems. Within a sample of 136 Im,
Sm and Blue Compact Dwarf galaxies, \cite{Hunter:2006} showed that 50
of them presented down-bending profiles, whereas 12 exhibited
up-bending ones. The ubiquity of multi-sloped profiles suggests that
they either form easily (perhaps through more than one mechanism)
and/or are very long-lived; otherwise only a small fraction of
galaxies would exhibit these features.

The down-bending profiles exhibit a characteristic U-shaped color
profile, both locally (\citealt{Bakos:2008}) and at high redshift
(\citealt{Azzollini:2008a}). The color gets bluer out to the break
radius, as one would expect from an inside-out formation scenario, and
then becomes redder past the break radius.  Numerical simulations by
\cite{Roskar:2008} and \cite{Sanchez-Blazquez:2009} have attributed
this excess of red emission in galactic outskirts to radial stellar
migration. According to these models, more than half of the old stars
currently found outside the break radius were actually born inside it,
and later migrated outwards. This scenario seems to be borne out by
observations of resolved stellar populations across the break radius
in lower-mass galaxies (\citealt{de-Jong:2007,Radburn-Smith:2011}) and
2D optical spectra (\citealt{Yoachim:2010,Yoachim:2012}).

Disk breaks have been detected up to $z \sim 1$
(\citealt{Perez:2004}), and studies have suggested that the break
radius increases with time (\citealt{Trujillo:2005};
\citealt{Azzollini:2008}). In principle, this suggests that these
features could be intimately linked to the inside-out growth of
galactic disks. However, the picture could well be different in barred
galaxies. Indeed simulations show that bars are expected to grow with
time (see \citealt{Athanassoula:2003} and references therein), so the
temporal growth of the break radius could be driven by that of the
bar, even in the absence of a significant inside-out evolution of the
disk itself.

One of the first explanations for the physical origin of disk breaks
appealed to the conservation of angular momentum of the infalling
material. \cite{van-der-Kruit:1987} showed that a collapsing gaseous
sphere settles onto a disk with a break corresponding to the maximum
specific angular momentum of the original spherical cloud. However, in
a more realistic scenario where gas is deposited in the outer parts
with varying angular momenta and timescales, the notion of a constant
cutoff radius becomes less likely.

Moreover, angular momentum can be subsequently redistributed if
non-axisymmetric features such as bars form, as shown by N-body
simulations (\citealt{Sellwood:1980, Athanassoula:2003}). Bars tend to
drive material within the corotation radius (CR) to smaller radii, and
material outside CR outwards, thus increasing the central stellar
density, while flattening the disk. Interestingly, though, bars
themselves can also give rise to radial breaks, as described by
\cite{Debattista:2006}. These simulations are quite demanding in terms
of the number of particles required to properly sample the outer disk,
and this comes at the expense of using a rigid halo rather than a live
one. In this regard, \cite{Foyle:2008} extended the study of
\cite{Debattista:2006} by using live halos and exploring a wider range
of galaxy properties. They found that the onset of breaks seems to be
governed by the ratio between the halo spin parameter $\lambda$ and
the disk mass fraction $m\sub{d}$. It has been also shown that a live
halo can absorb a substantial amount of angular momentum from the bar
(\citealt{Athanassoula:2002a}) $-$in fact much more than the tenuous
outer parts of the disk$-$ which will in turn affect the bar
properties, as well as the angular momentum absorbed by the outer
disk. A live halo might therefore modify the results of simulations
quantitatively, but perhaps not qualitatively.

If the break is linked to the bar then it will be inside its Outer
Lindblad Resonance (OLR). However, if the bar drives a spiral by
non-linear mode coupling (\citealt{Tagger:1987, Sygnet:1988}), then
the break should form inside the OLR of the spiral
(\citealt{Debattista:2006}), which is located further out than the bar
OLR, as the spiral pattern speed is lower than that of the
bar. Interestingly, recent simulations also show that under a
bar-spiral coupling, breaks can also form at the spiral CR
(\citealt{Minchev:2012}). Finally a break can come from
manifold-driven spirals, in which case its radius is not linked to the
outer resonance of the spiral, but is nevertheless located not far
from it (\citealt{Athanassoula:2012}).

From an observational point of view, E08 found that the break radius
in many down-bending profiles is located at 2 or 3 times the bar
radius, and thus proposed a possible connection with the OLR. This is
based on the fact that outer rings, which have been traditionally
associated with the OLR, have a radius around twice the bar radius
(\citealt{Kormendy:1979,Athanassoula:1982,Buta:1993}). The real
picture, however, can be substantially more complicated since the
precise location of resonances depends on the pattern speed of the bar
(and spiral arms if sufficiently massive) and the rotation curve of
the galaxy, both of which may change over time.

In contrast to the angular momentum framework, some studies favor star
formation thresholds as a likely culprit for disk breaks. If there is
a critical gas threshold for star formation
(\citealt{Kennicutt:1989}), then this may give rise to a break in the
radial profiles. However, the discovery of extended UV emission well
beyond the main optical disk of many galaxies
(\citealt{Gil-de-Paz:2005a, Thilker:2005, Zaritsky:2007}) challenges
this view. According to the survey by \cite{Thilker:2007}, in roughly
20\% of nearby galaxies this extended emission is in the form of
structured star-forming knots at extreme radii; other disks (around
10\%) present abnormally large and uniform areas with very blue UV-nIR
colors, although not so far from the main disk. A follow-up study by
\cite{Lemonias:2011}, using a larger sample and a slightly different
classification scheme, found the incidence of XUV-disks to be below
but close to 20\%.

The existence of star formation activity at such extreme radii in some
galaxies also implies that, at least in these cases, the break radius
might not necessarily correspond to the material with the maximum
angular momentum. As shown by \cite{Christlein:2008}, the rotation
curve probed by star-forming knots up to twice the optical radius
seems to be flat, meaning that this material has large angular
momentum. Star formation in the outskirts of disks could be triggered
by turbulent compression; together with other mechanisms driving star
formation in the inner parts, this might actually yield down-bending
profiles (\citealt{Elmegreen:2006}). Thus, depending on whether the
galaxy has an extended gas disk or not, and where the transition
between the inner and outer gas profiles takes place, the
superposition of these different mechanisms can lead to either
down- or up-bending profiles. In this context, XUV-disk host
galaxies are likely the ones with the shallowest outer gas
profiles. Even in these extreme cases a secondary outer truncation
might be present due to a star formation threshold associated with
either a sharp decrease in the gas density or decrease in the
efficiency of star formation mechanism at larger radii.

While resonances and/or star formation thresholds are normally invoked
to reproduce down-bending profiles, up-bending ones are often
explained through external mechanisms. \cite{Younger:2007} showed that
minor mergers can yield up-bending profiles: gas inflows towards the
center of the galaxy would steepen the inner profile, while the outer
one would expand as angular momentum is transferred outwards during
the interaction. Apart from minor mergers, \cite{Minchev:2012}
demonstrated that smooth gas accretion can also lead to a flat outer
disk, creating not only pure up-bending profiles (Type~III) but also
composite ones (Type~II + III).

Going back to down-bending profiles, amongst the different mechanisms
that can produce and/or modify this kind of breaks, bars should
receive special attention. Their non-axisymmetric potentials induce
radial transfer of gas, stars and angular momentum, which can lead to
a substantial rearrangement of the disk structure. Parameters like the
break radius or the ratio between the inner and outer disk
scale-lengths are therefore expected to be different in barred and
unbarred galaxies, and perhaps correlated with properties of the bar
such as its length or ellipticity.

There is consensus that the bar fraction in the local universe is
around $\sim 65$\% in the infrared (\citealt{Eskridge:2000,
  Knapen:2000, Whyte:2002, Menendez-Delmestre:2007,
  Marinova:2007}). Trends of the bar fraction with mass, color,
morphological type and environment have been also reported in the
literature (see, e.g., \citealt{Nair:2010,Masters:2011,Skibba:2012}
and references therein). Moreover, it has been shown that the redshift
evolution of the bar fraction is strongly dependent on the total
stellar mass of galaxies (\citealt{Sheth:2008fk}). The bar fraction
for massive disks is roughly constant up to $z \sim 0.8$, but it
declines considerably with increasing redshift for low mass
galaxies. Simulations show that the bar instability grows faster if
the disk is dynamically cold and/or rotationally supported
(\citealt{Athanassoula:1986, Sheth:2012}), and therefore this
differential evolution of the bar fraction has important implications
for the assembly of galactic disks. In particular, if bars are
responsible for disk breaks, and given that low-mass disks seem to
have acquired their bars only recently, it is worth investigating
whether this translates into systematic variations of the break
properties (radius, scale-length ratio, etc) as a function of stellar
mass.

Previous studies on disk breaks at low and high redshifts have relied
on rest frame optical images.  Probing galactic structure through
optical light profiles is hampered by the radial variations in dust
content, metallicity and mass-to-light ratio, all of which can make
the derived properties of disks significantly different from those of
the underlying old stellar population. In order to overcome these
limitations, here we rely on the Spitzer Survey of Stellar Structure
in Galaxies (\sfg\, \citealt{Sheth:2010}), a census of more than 2300
galaxies within 40\,Mpc imaged at 3.6\,$\micron$ and
4.5\,$\micron$. The \sfg\ dataset probes stellar surface densities as
low as 1\,M$_{\sun}$\,pc$^{-2}$, and therefore constitutes an ideal
benchmark to study the outskirts of galactic disks. Given the large
number of galaxies included in the survey, subsets of several hundreds
objects can be easily assembled after slicing the parent sample
according to different selection criteria, thus providing
unprecedented statistical power at these wavelengths. Here we present
a first analysis of disk breaks for a sample of more than 200 face on
disks, with stellar masses greater than $\sim 2 \times
10^9$\,M$_{\sun}$.

The paper is organized as follows. In Section~\ref{sec_sample} we
describe the selection criteria used to assemble our sample from the
parent \sfg\ survey. Section~\ref{analysis} deals with the technical
aspects of the analysis; in particular, we detail the data processing
and profile measurement (\S\ref{processing}), the classification of
disk profiles (\S\ref{classification}), and the measurement of the
properties of disks (\S\ref{disk_measurements}) and bars
(\S\ref{bar_measurements}). The main scientific results are discussed
in Section~\ref{results}, and we summarize our main conclusions in
Section~\ref{conclusions}.

\section{Sample selection}\label{sec_sample}
The full \sfg\ sample comprises of a total of more than 2300 nearby
galaxies. They were selected from the HyperLEDA database
(\citealt{Paturel:2003}) according to the following selection
criteria: radio-derived radial velocity
$v_{\mathrm{radio}}<3000$\,km/s (which corresponds to $d \lesssim
40$\,Mpc for $H_0 = 71$\,km\,s$^{-1}$\,Mpc$^{-1}$), Galactic latitude
$|b|>30^{\circ}$, total corrected blue magnitude
$m_{B_{\mathrm{corr}}}<15.5$ and blue light isophotal diameter
$D_{25}>1.0$\arcmin.

For the present work we started with the 827 galaxies that were
processed first through the \sfg\ pipelines (mosaic construction,
object masking and ellipse profile fitting, see
Section~\ref{processing}). On this subset of galaxies we applied three
selection cuts based on morphology, stellar mass and inclination. We
selected only disk-like galaxies ranging from S0's to Sd's (that is,
having numerical types $-3 \leq T \leq 7$), using the optical
morphological types compiled in HyperLEDA. Sdm and Sm galaxies, while
still exhibiting a spiral disk, usually present a patchy and
asymmetric morphology that complicates the measurement and
interpretation of radial profiles, so they were left out from our
sample.

We also decided to restrict our analysis to a well defined range in
stellar mass. Global properties of galaxies such as color, star
formation rate, stellar age, metallicity, gas fraction, etc., vary
with both the total stellar mass and the morphological type
(\citealt{Boselli:1997, Kauffmann:2003, Brinchmann:2004,
  Tremonti:2004, Schiminovich:2007}). However, the trends seem to be
tighter and better defined when plotted against stellar mass, hinting
that this is the main parameter that, at least to first order, governs
most of a galaxy's evolutionary path. Also, basing our analysis on the
stellar mass $-$or a reasonable luminosity proxy$-$ makes it easier to
compare our results with those at higher redshifts, where
morphological studies are challenging, because of both coarser
resolution and intrinsic departures from the classical Hubble types.

To that aim, here we use the 3.6\,$\micron$ luminosity as a stellar
mass tracer. It is worth noting that sources other than old stars
might contaminate the emission at this wavelength. \cite{Meidt:2012}
were able to isolate the old stellar light in a test sample of six \sfg\
galaxies, by applying an Independent Component Analysis to the
corresponding 3.6 and 4.5\,$\micron$ images. They concluded that hot
dust and Polycyclic Aromatic Hydrocarbons together may account for 5
to 15\% of the total integrated light at 3.6\,$\micron$ (see also
\citealt{Zibetti:2011}), while intermediate-age stars with low
mass-to-light ratios do not contribute more than 5\%. These values are
low enough so as not to compromise the usefulness of the
3.6\,$\micron$ luminosity as a stellar mass proxy.

Absolute magnitudes at 3.6\,$\micron$ were computed from the
asymptotic apparent magnitudes obtained with our pipeline (see
Section~\ref{processing}). Whenever possible, we relied on the mean
redshift-independent distances provided by the NASA Extragalactic
Database (NED). In the absence of these values, we estimated the
distance to each galaxy from its radial velocity $v_{\mathrm{radio}}$
and our adopted $H_0$. We retained only galaxies with
$M_{\mathrm{3.6\,\micron}}\ (\mathrm{AB}) < -18$\,mag. Using the
stellar mass-to-light ratio at 3.6\,$\micron$ derived by
\cite{Eskew:2012} (see Appendix~\ref{ML_ratio}), our magnitude cut
selects galaxies with stellar masses larger than $\sim 2 \times
10^9$\,M$_{\sun}$.

Finally, we required our galaxies to be at least moderately face-on,
having an axial ratio $b/a>0.5$ ($i<60^{\circ}$) at
$\mu_{3.6\,\micron} = 25.5$\,\maggie, which we take as our fiducial
surface brightness level for the outer parts of disks. In more
inclined galaxies, the shape of the isophotes can be strongly
distorted by the vertical structure of the disk, bulge and stellar
halo, thus precluding a straightforward interpretation of the light
profiles. Moreover, in order to investigate the role of bars in
shaping disk breaks, we need to obtain the deprojected values of
properties such as bar length or its ellipticity; by focusing on
galaxies close to face-on we minimize the impact of errors in the
assumed inclination angle.

After applying these three selection criteria we are left with 248
galaxies. After a visual inspection, we removed 30 galaxies that were
not suitable for our analysis. These mostly include highly-inclined
early-type disks such as the Sombrero galaxy, whose low axial ratios
are due to the spheroidal component. We also removed some highly
disturbed galaxies like NGC~0275, as well as galaxies whose radial
profiles are unreliable due to very bright foreground stars in the
field of view (such as NGC~6340).

\begin{figure}
\begin{center}
\resizebox{1\hsize}{!}{\includegraphics{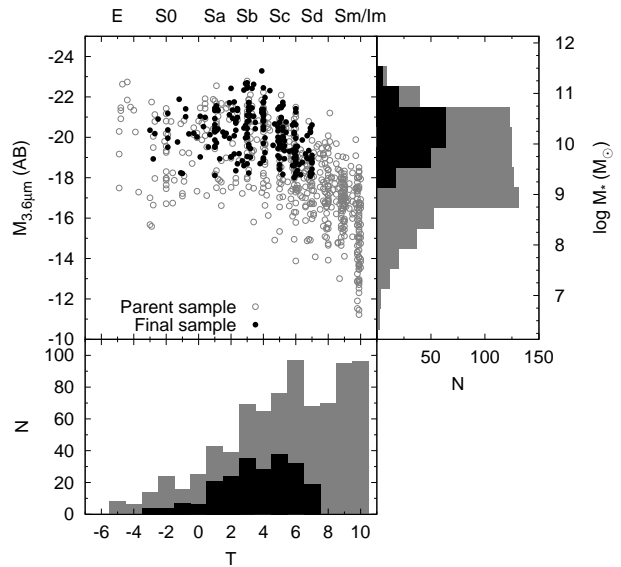}}
\caption{Distribution of Hubble types and absolute magnitudes at
  3.6\,$\micron$. The parent sample of 827 galaxies with fully reduced
  and processed data is shown in gray, whereas the final sample of
  218 face-on disks is shown in black.\label{sample}}
\end{center}
\end{figure}

Our final sample thus contains 218 galaxies, including both barred and
unbarred ones, whose main properties are quoted in
Table~\ref{sample_tab}. Figure~\ref{sample} shows the distribution of
morphological types and absolute 3.6\,$\micron$ magnitudes in both the
parent sample of 827 processed galaxies and the final sample. The
images and profiles of these galaxies, together with quantitative
measurements of their structural properties, are presented in
Appendix~\ref{atlas}.

\section{Analysis}\label{analysis}
\subsection{Data processing and profile measurement}\label{processing}
A description of the different parts comprising the \sfg\ pipeline is
presented by \cite{Sheth:2010}. Here we make use of the science-ready
images produced by our Pipeline~1. These images are flux-calibrated in
units of MJy\,sr$^{-1}$. The FWHM of the PSF is 1.7\arcsec, which maps
to a physical scale of 170\,pc at the median distance of our subsample
of 218 galaxies (21\,Mpc). Masking all relevant foreground stars and
background galaxies is essential in order to obtain reliable profiles
that reach low surface brightness levels. Object masks are produced by
our Pipeline 2 using SExtractor (\citealt{Bertin:1996}), and are later
visually checked and edited by hand, masking or unmasking pixels as
needed.

Here we describe Pipeline 3, which measures the sky level and noise,
and obtains radial profiles of surface brightness, position angle (PA)
and ellipticity ($\epsilon$). We measure the sky in two concentric and
adjacent elliptical rings around the galaxy. The major diameter of the
innermost ring is initially set by default to $2 \times D_{25}$, but
this value can be modified as needed, depending on the spatial extent
of each galaxy and the available background area within the field of
view. Each annulus is then azimuthally subdivided into 45 sectors which
serve as ``sky boxes''. These boxes are grown outwards, avoiding the
masked areas, until they contain 1000 unmasked pixels each. We then
compute the median sky value within each box. Should there be any
significant difference between the sky value in the inner and outer
annuli, we readjust the radii of the elliptical annuli
accordingly. The goal is to minimize the contamination from the galaxy
itself, while still ensuring that the derived background value is
representative of the area of the sky where the galaxy lies. In those
problematic cases where this method yields unreliable results, the sky
boxes are placed by hand.

Taking the galaxy coordinates from LEDA as input, we fit the
centroid of each source using the IRAF\footnote{IRAF is
  distributed by the National Optical Astronomy Observatories, which
  are operated by the Association of Universities for Research in
  Astronomy, Inc., under cooperative agreement with the National
  Science Foundation.} task {\sc imcentroid}. Using
these new and more accurate central coordinates, we run the task {\sc
  ellipse} to get radial profiles of surface brightness, ellipticity
and position angle, keeping the center fixed. We perform a first run
with a coarse radial step of 6\arcsec, and determine the ellipticity
and PA at a surface brightness level of 25.5 and 26.5\,\maggie. The
values at 25.5\,\maggie\ are typically robust enough against
variations in, e.g$.$, the sky level or the degree of masking, so we
adopt them as representative of the shape of the outer parts of our
galaxies. A second {\sc ellipse} run is then performed, keeping the
ellipticity and PA fixed to these outer values, and using a finer
radial increment of 2\arcsec\ that better matches the FWHM of the
PSF. This is the fit that we use to measure disk breaks. A third
fit with a step of 2\arcsec\ but free ellipticity and PA is also
performed; we use this third fit to determine the properties of bars
(length, ellipticity and PA).

The uncertainty in the surface brightness is computed following the
methodology described in \cite{Gil-de-Paz:2005} and
\cite{Munoz-Mateos:2009a}. There are two main sources of error at each
radius: Poisson noise in the incoming flux and errors in the
determination of the sky level. The former is derived by replicating
the ellipse measurements on the IRAC coverage maps, to take into
account that the effective gain might vary spatially, depending on the
mosaicking pattern. The uncertainty in the sky level, on the other
hand, comprises pixel-to-pixel noise as well as large-scale background
variations. These values are computed from the rms within and among
the different sky boxes, respectively. Large-scale variations
constitute the dominant source of error in the outskirts of our
galaxies (see also \citealt{Comeron:2011a} and
\citealt{Martin-Navarro:2012}).

An additional correction needs to be applied to the surface
photometry, in order to account for both the extended emission of the
PSF and the scattered light on the detector. According to the IRAC
handbook\footnote{\tt
  http://ssc.spitzer.caltech.edu/irac/calib/extcal/}, if
$F_{\mathrm{obs}}$ is the total flux {\it inside} an elliptical
aperture with major and minor radii $a$ and $b$, then the
aperture-corrected flux can be obtained as:
\begin{equation}
F_{\mathrm{corr}} (r_{\mathrm{eq}}) = F_{\mathrm{obs}} (r_{\mathrm{eq}}) \times (A e^{-r_{\mathrm{eq}}^B}+C)\label{eq_F_aper_corr}
\end{equation}
where $r_{\mathrm{eq}}=\sqrt{ab}$ is the equivalent radius of the
aperture in arcseconds, and the constants $A$, $B$ and $C$ are equal
to 0.82, 0.370 and 0.910, respectively, for the 3.6\,$\micron$ band.
The uncertainty in this correction is estimated to lie within 5\% to
10\%.

Again, this expression is only valid for the total flux {\it inside}
an aperture. In order to obtain an analogous expression for the
surface brightness {\it along} a given isophote, $I_{\mathrm{corr}}$,
we can simply apply a series expansion to Eq.~\ref{eq_F_aper_corr}:
\begin{eqnarray}
I_{\mathrm{corr}} (r_{\mathrm{eq}}) &=& I_{\mathrm{obs}} (r_{\mathrm{eq}}) \times (A e^{-r_{\mathrm{eq}}^B}+C)- \nonumber \\
&&{}-ABr_{\mathrm{eq}}^{B-2}e^{-r_{\mathrm{eq}}^B} F_{\mathrm{obs}} (r_{\mathrm{eq}})/(2 \pi) \label{eq_I_aper_corr}
\end{eqnarray}

Note that for large apertures, $F_{\mathrm{corr}} \simeq C \times
F_{\mathrm{obs}}$ and $I_{\mathrm{corr}} \simeq C \times
I_{\mathrm{obs}}$, but here we explicitly use
Eqs.~\ref{eq_F_aper_corr} and \ref{eq_I_aper_corr} at each radius.

The surface brightness can be then computed as:
\begin{eqnarray}
\mu_{\mathrm{corr}} (\mathrm{AB\,mag\,arcsec}^{-2}) &=& -2.5 \log
[I_{\mathrm{corr}} (\mathrm{MJy\,str}^{-1})]+ \nonumber \\
&&{}+21.097
\end{eqnarray}
where the zero-point has been computed according to the standard
definition of the AB magnitude scale (\citealt{Oke:1974}).

As a byproduct of the surface photometry, we also obtain the
asymptotic magnitude for each galaxy, that is, the magnitude that we
would measure with a hypothetically infinite aperture. We do so by
calculating $m(r)$, the total magnitude within a radius $r$, as a
function of the local magnitude gradient, $dm(r)/dr$. In the outer
parts of galaxies, these variables are linearly related; we therefore
apply a linear fit and take the y-intercept $-$the magnitude at a null
gradient$-$ as our asymptotic magnitude.

\subsection{Classifying disk profiles}\label{classification}
Following the scheme laid out by previous work (see, e.g.,
\citealt{Pohlen:2006}, E08 and references therein), we divide our
profiles into three main broad categories: Type~I (single
exponential), Type~II (down-bending) and Type~III (upbending). In
Fig.~\ref{sampleprof} we show individual examples of each of these
types. Type~I profiles require no further explanation; more details on
Type~II and III profiles are given below.

\subsubsection{Type~II profiles}
\begin{figure}
\begin{center}
\resizebox{1\hsize}{!}{\includegraphics{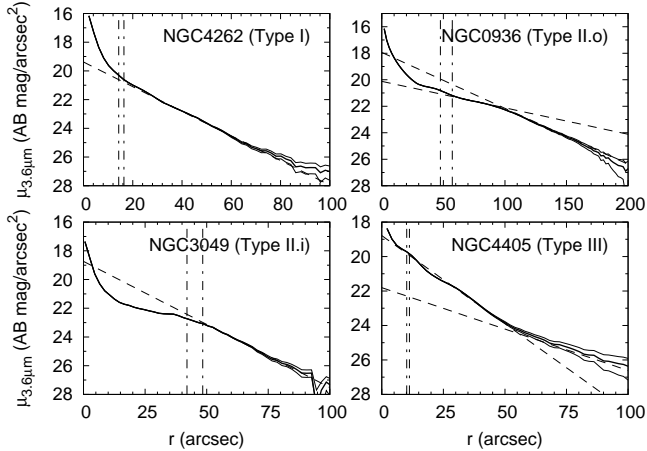}}
\caption{Sample profiles showing the different morphologies of our
  profiles: single exponential (Type~I), down-bending (Type~II) and
  up-bending (Type~III). The vertical dot-dashed lines show the lower
  and upper limits to the bar size (Sect.~\ref{bar_measurements}), but
  these three types are present in unbarred galaxies as well. In
  Type~II barred galaxies, the break can be at or inside the bar
  radius (Type~II.i) or outside it (Type~II.o). Dashed lines show
  fits to each component of the disk profile.\label{sampleprof}}
\end{center}
\end{figure}

These profiles are characterized by a break beyond which the profile
becomes steeper. In other words: the radial scale-length of the outer
disk is shorter than that of the inner one. In most cases, the break
lies in the outer parts of the disk and, in particular, outside the
radius of the bar, should there be one. Following E08, we refer to
these profiles as Type~II.o (``outer''). On the other hand, some
barred galaxies exhibit a break which is so close to the bar radius
that the profile actually looks purely exponential. However, when
extrapolating this exponential towards the central regions, the result
overpredicts the actual surface brightness of the bar and/or the bulge
(Fig.~\ref{sampleprof}). This is in contrast to genuine Type~I
profiles, where the extrapolated exponential always lies below the bar
and bulge components. To distinguish between these two cases, these
``inner'' breaks are denoted as Type~II.i.

E08 went one step further and subdivided Type~II.o profiles of barred
galaxies into two categories, depending on their presumed physical
origin. When the breaks were found between 2 or 3 times the bar
radius, they referred to them as Type~II.o-OLR, since they could be
potentially linked to the Outer Linblad Resonance. If, however, the
breaks were further out, they labelled them as Type~II.o-CT profiles,
since they seemed to be more similar to the ``Classical Truncations''
found in unbarred galaxies. While physically motivated, here we prefer
to defer these interpretations to the Analysis section (in particular,
Section~\ref{resonances}), where a more quantitative comparison
between the properties of breaks and bars will be presented.

\subsubsection{Type~III profiles}
Unlike Type~II profiles, where the light distribution bends down
beyond the break radius, Type~III profiles exhibit a flatter slope
outside the break. It should be noted, however, that the stellar
haloes of early-type disks might contribute significantly to the light
emitted in the outer parts of these galaxies, and these should be
distinguished from the ones caused by a shallower scale-length of the
outer disk itself. Besides, \cite{Comeron:2012} showed that the thick
disk component with a flatter scale-length than the thin disk can also
lead to up-bending profiles.

In moderately inclined galaxies, the ellipticity profile is of great
help here: if the ellipticity drops in the outer parts, the upbending
is most likely due to the spheroidal component, which is rounder than
the disk seen in projection.  Following E08, we denote those profiles
as Type~III-s (``spheroidal''). On the contrary, if the ellipticity
remains roughly the same beyond the break radius, then we are probably
witnessing a change of slope of the disk itself, and therefore name
these profiles as Type~III-d (``disk'').

This method cannot be applied in galaxies close to face-on, where the
ellipticity remains low throughout the whole profile. However, if we
see structured emission in the outer parts of these galaxies, we
consider these profiles to be Type~III-d as well. Also, besides the
ellipticity signature, the breaks in Type~III-d profiles are usually
sharper and better defined than in Type~III-s ones, where the
transition between the inner and outer slope is smoother and more
gradual. 

A detailed analysis of Type~III profiles is left for future papers
but, for completeness, the galaxies in our sample exhibiting this kind
of profile are also quoted in Table~\ref{prof_measurements}.

\subsection{Measuring the properties of disk breaks}\label{disk_measurements}
In order to characterize the properties of our disk profiles we follow
a methodology similar to that established by previous authors in the
field. As described before, for this particular purpose we employ the
profiles measured with fixed ellipticity and position angle and a
2\arcsec\ radial increment. The simplest case is that of a Type~I
profile, where we apply a linear fit to the disk-dominated region:
\begin{equation}
I(r)=I_0 e^{-r/h} \Rightarrow \mu (r) = \mu_0 + 1.086 \frac{r}{h}
\end{equation}

where $\mu_0$ is the central surface brightness of the disk and $h$
its exponential scale-length. The inner boundary of the fitted region
is set to exclude the bulge, as well as the shoulder of the bar,
should there be one. The outer boundary is tuned according to the
uncertainty in the surface brightness, to prevent spurious features
and/or errors in the sky subtraction from biasing the fit. In
practice, for each galaxy we compute a critical surface brightness
value $\mu_{\mathrm{crit}}$ beyond which the uncertainty in the light
profile exceeds 0.2\,mag\,arcsec$^{-2}$. We typically place the outer
boundary around the radius where this surface brightness is reached.

Type~II and III profiles require a more sophisticated approach. In
order to simultaneously fit the inner and outer disks, we rely on the
broken exponential function proposed by E08:
\begin{equation}
I(r) = S I_0 e^{-\frac{r}{h_i}}\left[ 1 + e^{\alpha(r-R_{\mathrm{br}})} \right]^{\frac{1}{\alpha}\left(\frac{1}{h_i}-\frac{1}{h_o}\right)}\label{eq_broken_exponential}
\end{equation}

Here, $h_i$ and $h_o$ are the scale-lengths of the inner and outer
disks, respectively, $I_0$ is the central intensity of the inner disk
and $R_{\mathrm{br}}$ is the break radius. The coefficient $\alpha$
determines the sharpness of the break, with low values corresponding
to a smooth transition between the inner and outer disk, and high
values yielding a sharper break. The scaling factor $S$ is defined
as\footnote{Note that the definition of $S$ in E08 (their Eq.~6)
  contains a small typo: the order of the $1/h_i$ and $1/h_o$ terms
  should be reversed.}:
\begin{equation}
S=\left(1+e^{-\alpha R_{\mathrm{br}}} \right) ^{\frac{1}{\alpha}\left(\frac{1}{h_o}-\frac{1}{h_i}\right)}
\end{equation}

\begin{figure}
\begin{center}
\resizebox{1\hsize}{!}{\includegraphics{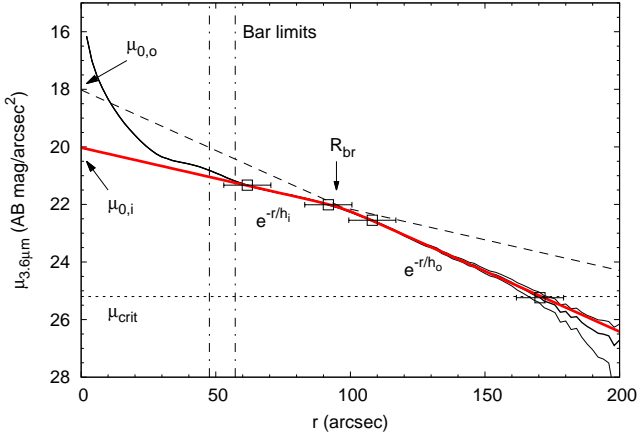}}
\caption{Sample fit of the broken exponential function to the light
  profile of NGC~0936, showing the nomenclature used throughout the
  paper. The profile itself is shown as a thick black line, whereas
  the thinner lines above and below it show the $\pm 1\sigma$
  uncertainty. The four squares delimit the inner and outer portions
  of the disk, from which initial guesses of the slopes and central
  surface brightnesses are obtained. Their error bars delimit the
  range used in the Monte Carlo simulations. The best-fitting broken
  exponential described by Eq.~\ref{eq_broken_exponential} appears in
  red. Even though this red curve is plotted spanning the whole radial
  range, the fit itself is done only between the first and fourth
  square points. Dashed lines show extrapolations of the inner and
  outer disks. The dotted line marks the critical surface brightness
  level $\mu_{\mathrm{crit}}$, beyond which $\Delta \mu >
  0.2$\,mag\,arcsec$^{-2}$. Finally, vertical dotted-dashed lines mark the
  minimum and maximum of our estimates of the bar radius
  (Sect.~\ref{bar_measurements}).\label{sample_fit}}
\end{center}
\end{figure}

An example of such a fit is shown in Fig.~\ref{sample_fit} for
NGC~0936. We first identify the inner and outer portions of the disk
profile, and delimit them with two points each (shown here as small
squares). Again, the innermost limit of all is set to avoid
contamination from the bulge and bar, whereas the outermost one is
placed in general around $\mu_{\mathrm{crit}}$. We then apply
individual linear fits to each portion of the profile separately. From
these we get initial estimates for both slopes and y-intercepts, as
well as for the break radius (from the intersection of both fits). The
resulting values are then used as input guesses for the parameters in
Eq.~\ref{eq_broken_exponential}, which is fitted iteratively to all
data-points bracketed by the innermost and outermost boundaries, using
a Levenberg-Marquardt algorithm.

The final values of the fitted parameters might depend on our
particular placement of the four range delimiters, as well as on the
accuracy of the sky subtraction. In order to get a handle on the
uncertainties that these factors might introduce, we proceed in a
Monte Carlo fashion. For a given profile we generate 1000 simulations,
in each one of which both the range delimiters and the sky value are
randomly modified at the same time. The delimiters are shuffled around
their central positions following a uniform probability distribution,
with a half-width equal to $\pm 0.05 \times R_{25.5}$. This interval
is shown in Fig.~\ref{sample_fit} as horizontal error bars. In some
cases this might place the innermost delimiter somewhat inside the
bulge- or bar-dominated region of the profile, but we do actually want
to include this in the error budget. Also, note that by using a
fractional value of the total disk size, we allow for a proportionally
larger margin of error in galaxies with larger apparent sizes. As for
the sky level, we draw random values from a gaussian distribution
whose standard deviation matches the large-scale background $rms$
measured in the image (as explained in Section~\ref{processing}),
which is the main source of uncertainty in $\mu$ at large radii. After
automatically fitting each simulated profile, we end up with a set of
1000 values for each one of the structural parameters. Out of these
distributions we get the corresponding upper and lower $1\sigma$
uncertainties, which are quoted in Table~\ref{prof_measurements}.

\subsection{Measuring bar properties}\label{bar_measurements}
The potential impact of a bar in shaping a galactic disk might depend,
at least to first order, on its length and its strength. Longer bars
have access to more material within the disk, and can therefore be
responsible of galaxy-wide radial rearrangement of both stars and gas
(that in turn might eventually be converted into new stars). Bar
strength, on the other hand, measures the non-axisymmetric torque
exerted by the bar. It is usually defined as the ratio between the
maximum tangential force normalized by the mean axisymmetric radial
force inside (\citealt{Combes:1981}). Obtaining the bar strength
directly from this definition requires a detailed evaluation of the
gravitational potential of the galaxy. This can be done from a
galaxy's IR image, provided that some assumptions on the mass-to-light
ratio and disk scale-height are made (see, e.g.,
\citealt{Quillen:1994,Buta:2001, Laurikainen:2002}). Such an analysis
is beyond the scope of this paper; however, it has been shown that the
maximum bar ellipticity constitutes a reasonable proxy for the bar
strength (\citealt{Laurikainen:2002a, Comeron:2010}). In general, bars
with larger ellipticities tend to be stronger and exert greater
torques.

Several methods and criteria have been proposed by different authors
to identify bars and measure their structural properties (length,
ellipticiy and PA). Among the most widely used techniques are those
based on ellipse fitting (\citealt{Wozniak:1995, Knapen:2000,
  Athanassoula:2002, Laine:2002, Sheth:2000, Sheth:2002, Erwin:2005a,
  Gadotti:2007, Menendez-Delmestre:2007}). Other methodologies have
also been used extensively, such as direct visual measurements
(\citealt{Kormendy:1979, Martin:1995, Hoyle:2011}), 2D image
decompositions (\citealt{Prieto:2001, Peng:2002, de-Souza:2004,
  Laurikainen:2005, Reese:2007, Gadotti:2008}; Kim et al 2012$.$, in
prep.), Fourier analysis (\citealt{Ohta:1990, Elmegreen:1985,
  Buta:2006, Athanassoula:2002, Laurikainen:2007}) and cuts along the
bar major axis (\citealt{Athanassoula:2002, Elmegreen:1985}).

\begin{figure*}
\begin{center}
\begin{tabular}{cc}
\resizebox{0.3\hsize}{!}{\includegraphics{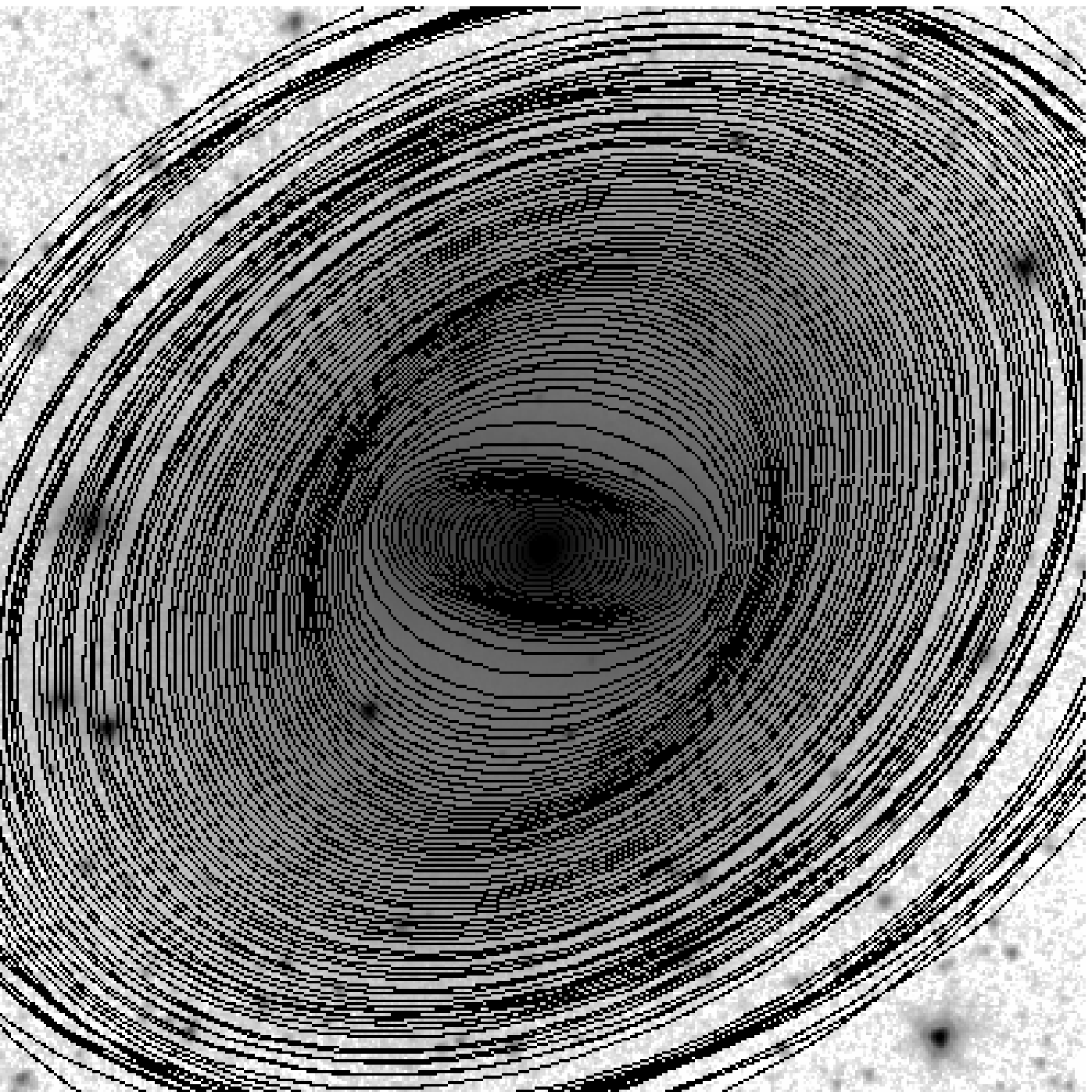}} & \multirow{2}{*}[44ex]{\resizebox{0.52\hsize}{!}{\includegraphics{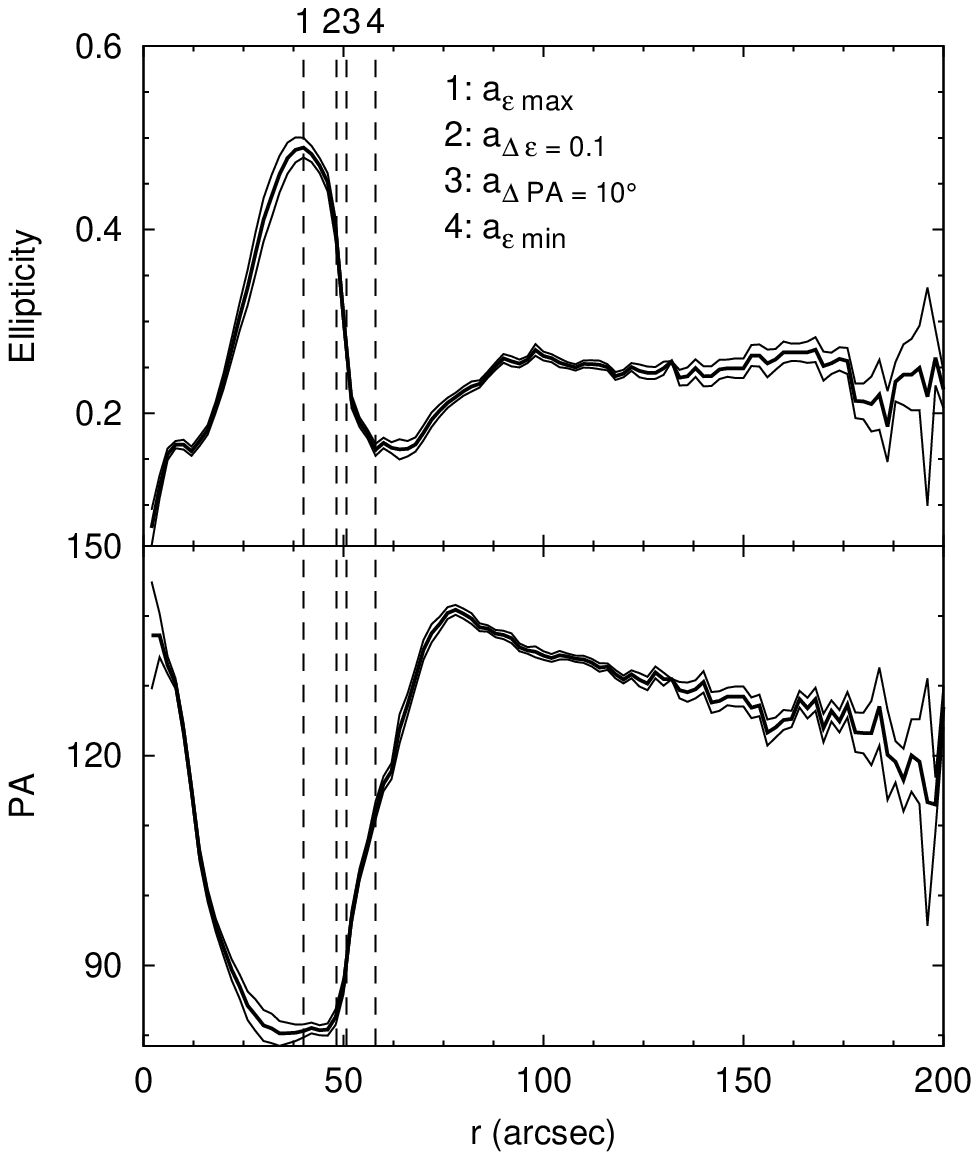}}}\\
\resizebox{0.3\hsize}{!}{\includegraphics{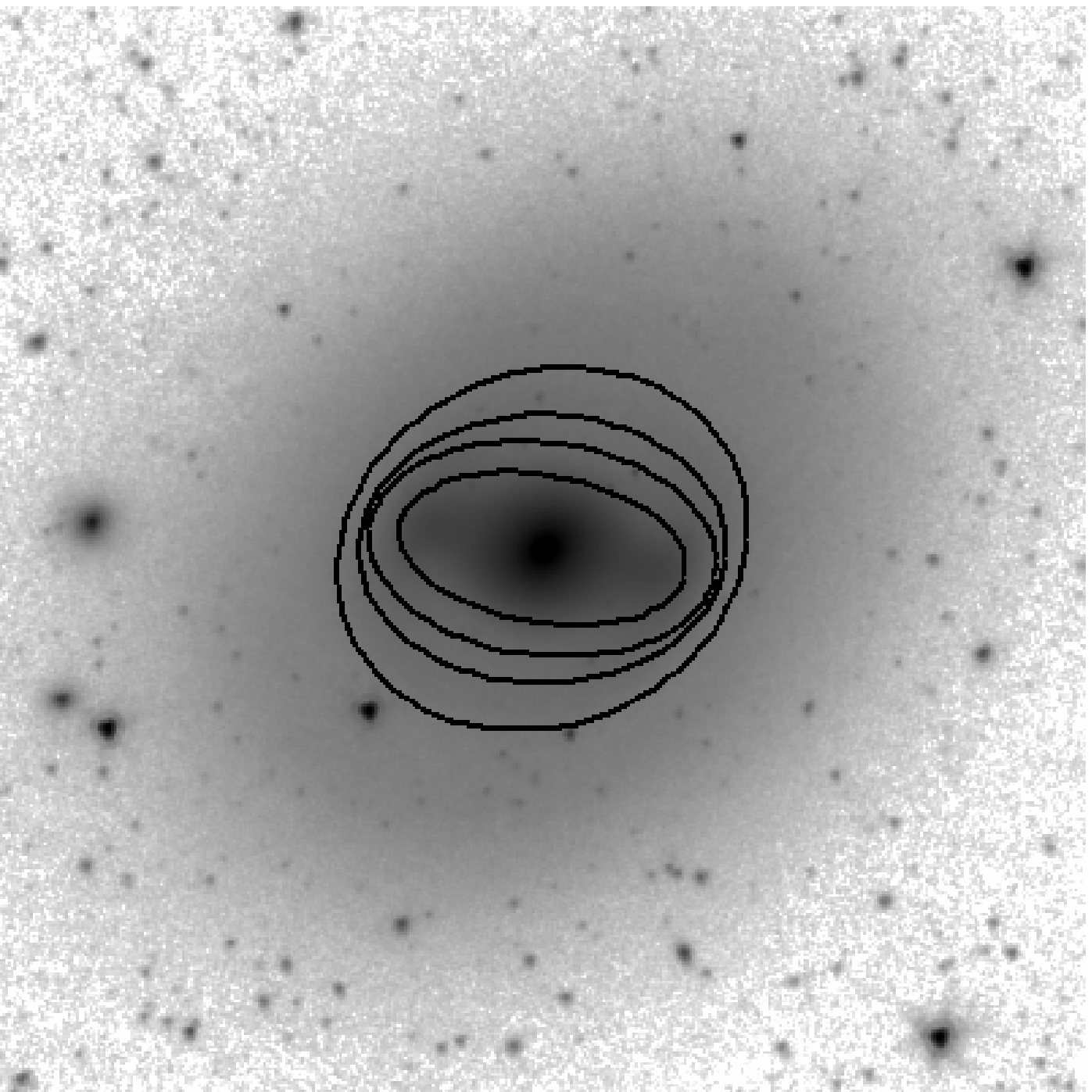}} & \\
\end{tabular}
\caption{Signature left by a bar in the ellipticity and position angle
  profiles. The top-left panel shows the 3.6\,$\micron$ image of
  NGC~0936 with the ellipse fitting results overlaid with a spacing of
  2\arcsec\ along the major axis between adjacent isophotes. The
  ellipticity and position angle profiles are displayed to the right;
  the thin lines above and below each profile show the $\pm 1\sigma$
  fitting error. The four estimations of the bar radius described in
  the text are marked with vertical dashed lines, and the
  corresponding four ellipses are overlaid in the bottom-left
  panel.\label{sample_bar}}
\end{center}
\end{figure*}

Here we rely on our radial profiles to determine the structural
properties of bars. Figure~\ref{sample_bar} depicts the typical
signature left by a bar on the ellipticity and PA profiles. Within the
bar region, the ellipticity increases monotonically and then decreases
by $\Delta \epsilon > 0.1$ as the isophotes begin to probe the
disk. The position angle, on the other hand, remains roughly constant
along the bar (except perhaps at the center), and then changes by
$\Delta PA \gtrsim 10^\circ$ at the bar ends. How abrupt these changes
in $\epsilon$ and PA are depend on how the spiral arms merge with the
bar ends, as well as on the orientation of the bar relative to the
projected disk (see \citealt{Menendez-Delmestre:2007} for examples and
a detailed discussion).

Bars in our sample were identified by simultaneously inspecting the
images and profiles for each galaxy, looking for the signatures
described above. Those cases where the presence of a bar was dubious
were tagged as ``candidate bars''. These usually correspond to cases
where the profiles exhibit a signature reminiscent of that of a bar,
but where the image does not convincingly confirm its presence. In
particular, some late-type disks exhibit elongated structures that
may result from a chance alignment of a few bright HII regions. Also,
non-axisymmetric bulges may be present in many galaxies
(e.g. \citealt{Zaritsky:1986})

After several trials and experiments, and building on previous work,
we decided to settle on four different measurements of the bar radius
based on the ellipticity and position angle profiles:

\begin{enumerate}
\item $a_{\epsilon\ max}$, the radius where the ellipticity of the bar is maximum.
\item $a_{\epsilon\ min}$, the radius where there is a local minimum in ellipticity after the previous maximum.
\item $a_{\Delta\epsilon=0.1}$, the radius where the ellipticity drops by 0.1 with respect to the maximum one.
\item $a_{\Delta PA=10^\circ}$, the radius where the position angle differs by 10$^\circ$ from the one at $a_{\epsilon\ max}$.
\end{enumerate}

These four measurements are shown in Fig.~\ref{sample_bar}. Some
observations (e.g$.$ \citealt{Wozniak:1995}) and N-body simulations
(e.g$.$ \citealt{Athanassoula:2002}) suggest that 
$a_{\epsilon\ max}$ underestimates the true radius of the
bar. We therefore adopt $a_{\epsilon\ max}$ as a lower limit for the bar
radius. As for the upper limit, we take whichever of the three
remaining measurements is smallest\footnote{Note that, in principle,
  it would not be necessary to compute $a_{\epsilon\ min}$, since
  $a_{\Delta\epsilon=0.1}$ will generally be smaller, almost by
  definition. However, we did find a few galaxies for which the
  ellipticity drop at the end of the bar was smaller than (but very
  close to) 0.1. In these few cases, $a_{\epsilon\ min}$ was found to
  be a much better measure of the bar radius.}. On average, after
deprojection (see below) the upper limits are 20\% larger than the
lower ones, in agreement with \cite{Erwin:2005}. Our final bar radius
for each galaxy is obtained by averaging the corresponding lower and
upper limits. The bar ellipticity and PA are assumed to be those at
$a_{\epsilon\ max}$.

All these three bar properties $-$radius, $\epsilon$ and PA$-$ need to
be corrected for inclination to obtain the intrinsic face-on
values. To do so, we follow the 2D de-projection method described in
\cite{Gadotti:2007}, which yields the true major and minor axes and
position angle of an ellipse seen in projection. This constitutes an
improvement over simpler methods whereby bars are assumed to be
one-dimensional lines. According to \cite{Gadotti:2007}, the
geometrical parameters resulting from this analytical de-projection
agree well with those measured on de-projected images.

This method requires knowing the line of nodes of the projected disk
and its inclination angle. We derive these values from the PA and
$\epsilon$ at 25.5\,\maggie, assuming that the outer disk at this
surface brightness level is intrinsically circular. Note that
kinematic PA and inclinations do not exist for such a large sample of
galaxies, so this photometric approach is a simple yet necessary
workaround in order to treat the whole sample homogeneously.

The intrinsic thickness of the disk is not expected to represent an
important issue here, either, because most of our galaxies are just
moderately inclined ($b/a>0.5$). If a galaxy is described as an oblate
spheroid (\citealt{Hubble:1926}), then the ``true'' inclination angle
$i$ can be estimated as:
\begin{equation}
\cos^2 i = (q^2-q_0^2) (1-q_0^2)^{-1}\label{eq_inclination}
\end{equation}
where $q=b/a$ is the observed axial ratio and $q_0$ is the intrinsic
one when the galaxy is viewed edge-on. Values of $q_0 \simeq 0.2$ are
usually assumed in the near-IR (e.g$.$ \citealt{Courteau:2007} in the
2MASS bands) and the mid-IR (e.g.$.$ \citealt{Comeron:2011} at
3.6\,$\micron$). For samples with $b/a>0.5$ such as ours, disk
thickness should not bias the assumed inclination angles by more than
$\sim 2^\circ$.

\section{Results}\label{results}

\subsection{Global properties of down-bending profiles}
\begin{figure}
\begin{center}
\resizebox{1\hsize}{!}{\includegraphics{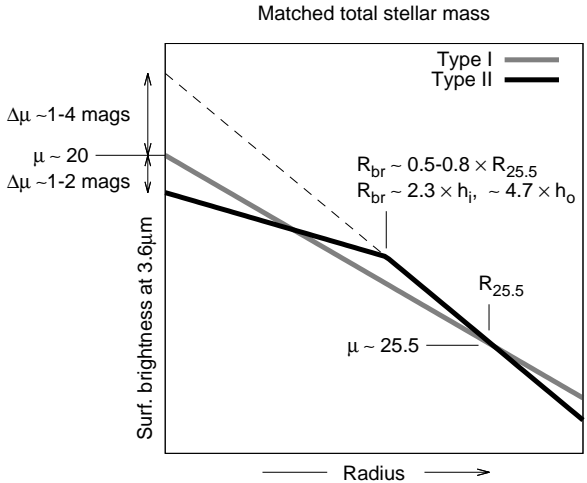}}
\caption{Schematic depiction of the structural properties of a
  down-bending disk, relative to a single-exponential one in a galaxy
  with the same total stellar mass. This plots summarizes the results
  of Figs.~\ref{T2_h_mu0}, \ref{T2_Rdisk_Mirac1} and
  \ref{T2_Rbreak_Mirac1}. We show the break radius $R_{\mathrm{br}}$
  relative to the isophotal radius $R_{25.5}$ and the inner and outer
  scale-lengths, $h_i$ and $h_o$. We also show the typical
  extrapolated central surface brightness of the inner and outer disk,
  relative to that of a single-exponential
  profile.\label{T1_T2_scheme}}
\end{center}
\end{figure}

Before diving into an in-depth analysis of the effect of bars on light
profiles, it is illustrative to first gauge the overall
characteristics of each type of profile, regardless of the presence or
absence of a bar. In the following subsections we describe the global
distribution of disk scale-length, surface brightness, disk isophotal
size, break radius and stellar density at the break. As a guide, the
main results are schematically summarized in Fig.~\ref{T1_T2_scheme},
which presents a simple cartoon comparison of a single-exponential
profile and a down-bending one for galaxies of the same total stellar
mass.

\subsubsection{Disk scale-length}
\begin{figure*}
\begin{center}
\resizebox{0.49\hsize}{!}{\includegraphics{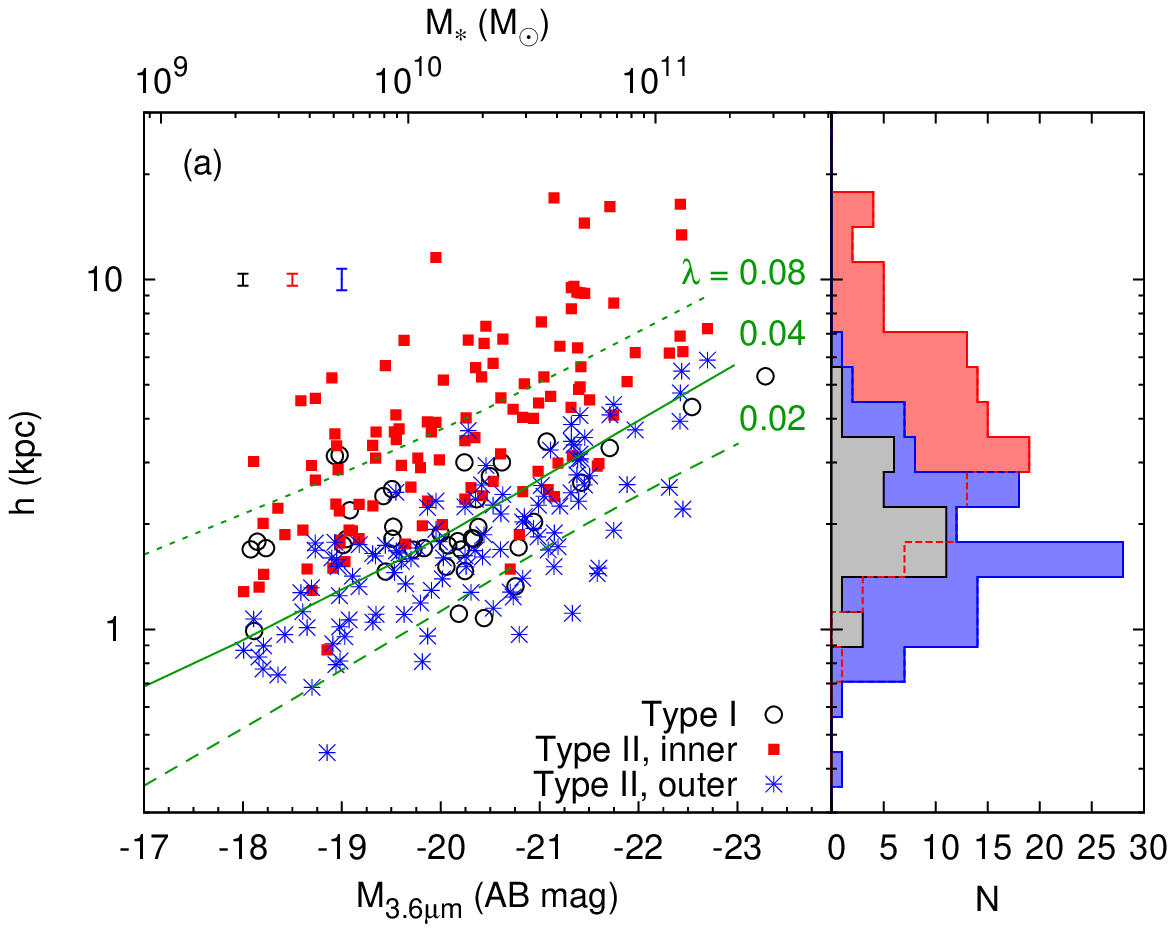}}
\resizebox{0.49\hsize}{!}{\includegraphics{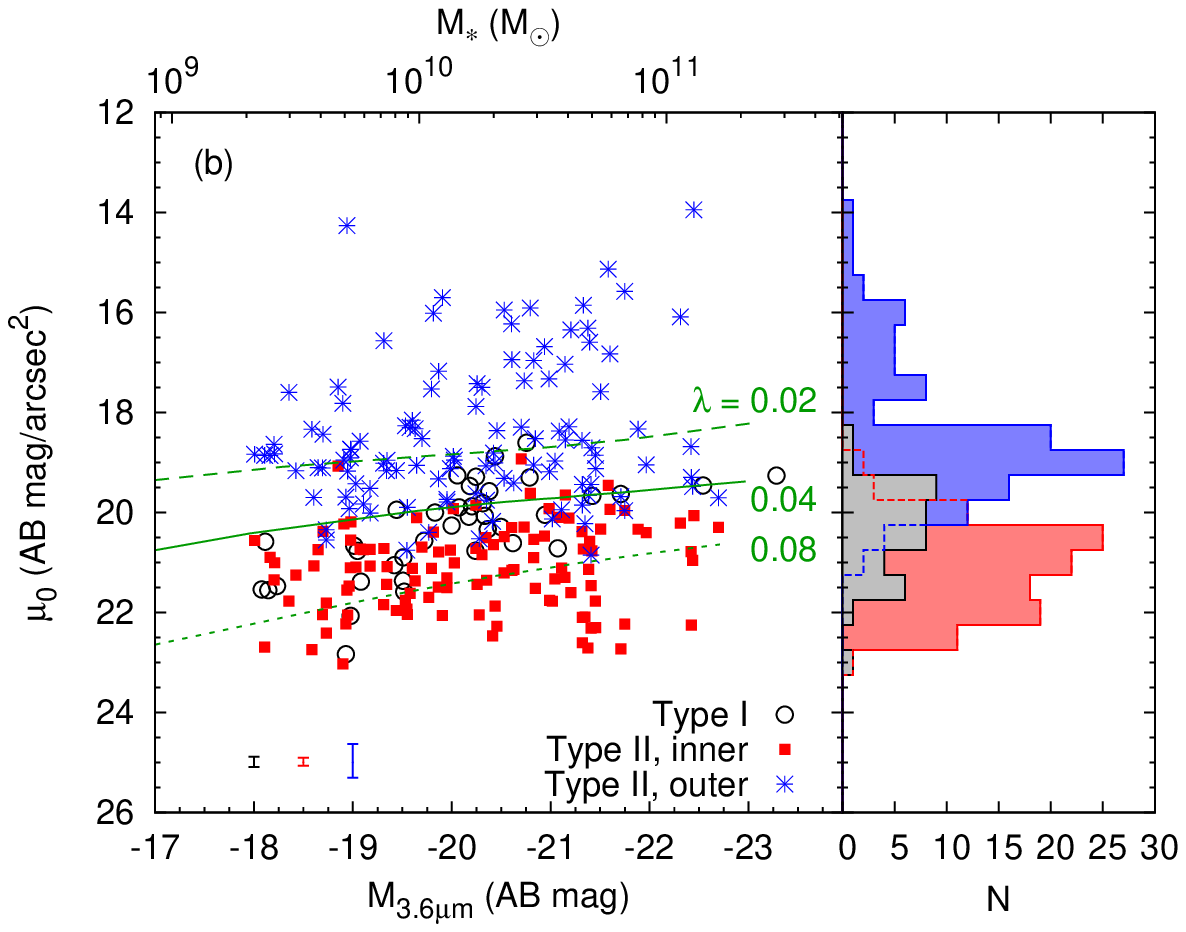}}\\
\caption{Distribution of disk scale-lengths (a) and extrapolated
  central surface brightness (b) as a function of absolute magnitude
  at 3.6\,$\micron$. Single-exponential profiles are marked with open
  symbols. For down-bending profiles, red squares and blue diamonds
  correspond to the inner and outer disk, respectively. The values of
  $\mu_0$ have been corrected for inclination. Median error bars are
  shown with the corresponding color. The green lines show the
  predictions of the models of \cite{Boissier:2000} (mostly valid for
  single-exponential disks) for selected values of the spin parameter
  $\lambda$. The histograms to the right show the distribution of the
  corresponding parameters.\label{T2_h_mu0}}
\end{center}
\end{figure*}

In Fig.~\ref{T2_h_mu0} we plot the disk scale-length and central
surface brightness of Type~II profiles (for both the inner and outer
disk) as a function of the absolute 3.6\,$\micron$ magnitude. As a
reference, we compare these properties with those of Type~I
profiles. This allows us to establish a parallelism between
single-exponential and down-bending profiles in galaxies with the same
total stellar mass\footnote{Note that the total stellar mass includes
  the contribution of the bulge, should there be one.}. The histograms
to the right show the overall distribution of these structural
parameters. In the particular case of the disk scale-length, our
measurements are consistent with the grid of simulations of disk
breaks by \cite{Foyle:2008}, who found inner disk-scalengths in the
range 1-10\,kpc, and outer disk scale-lengths between 0.5-5\,kpc.

Panel (a) shows a clear correlation between disk scale-length and
absolute magnitude. Such a trend between galaxy luminosity (or mass)
and size (as measured by either disk scale-length or effective radius)
has been already extensively analyzed at optical and near-IR
wavelengths (\citealt{Courteau:1999, Shen:2003, MacArthur:2004,
  Barden:2005, Trujillo:2006, Courteau:2007, Gadotti:2009,
  Dutton:2011}). It constitutes one of the basic scaling laws of
galactic disks (\citealt{Mo:1998, Boissier:2000, Firmani:2000,
  Firmani:2009, Brook:2006, Dutton:2007, Dutton:2011, Somerville:2008,
  Brooks:2011}). Observations have demonstrated that the scatter in
scaling laws involving disk scale-length is usually larger than in
other empirical laws such as the Tully-Fisher relation
(\citealt{Tully:1977}). There are several factors that may account for
this:

\begin{enumerate}
\item In the particular case of $h$ versus circular velocity, the
  latter is {\it not} usually measured at $h$ but further out, in the
  dark matter dominated regime.
\item The slope of the trend between disk scale-length and luminosity
  seems to vary systematically along the Hubble sequence
  (\citealt{Courteau:2007}).
\item The disk scale-length can change with time due to secular
  processes, such as inside-out growth of the disk or radial stellar
  migration. This can blur the connection between $h$ and the
  dynamical properties of the halo.
\item The scale-length of a disk does not only depend on the total
  mass, but also on the dimensionless spin parameter $\lambda$ (see
  below).
\item Since most disks have more than one exponential, the notion of a
  single scale-length is simply ill defined (see below).
\end{enumerate}

Regarding issue 4, the spin parameter $\lambda$
(\citealt{Peebles:1969}) relates a system's angular momentum, its
binding energy and its mass, and measures the degree of rotational
support. In general, for a given total mass, galaxies with higher
values of $\lambda$ exhibit flatter disks (\citealt{Dalcanton:1997,
  Boissier:2000}). The distribution of $\lambda$ usually follows a
log-normal function, both in simulations (\citealt{Barnes:1987,
  Warren:1992, Bullock:2001, Gardner:2001, Vitvitska:2002}) and
observations (\citealt{Syer:1999, Hernandez:2007, Berta:2008,
  Cervantes-Sodi:2008, Munoz-Mateos:2011}). The distribution peaks
around $\lambda \sim 0.03-0.04$, with a spread that can partly account
for the observed scatter in disk scale-lengths for a fixed mass.

To illustrate this, in Fig.~\ref{T2_h_mu0} we overlay the trends
predicted by the models of \cite{Boissier:2000}. Using the
$\Lambda$-CDM scaling laws as a working framework, these models
incorporate analytical yet physically motivated prescriptions for
radially-varying gas accretion, star formation and chemical
enrichment. For any pair of values of the spin $\lambda$ and the
circular velocity $V_\mathrm{C}$ of the parent dark matter halo, the
models yield the temporal evolution of the radial profiles at
different wavelengths. Figure~\ref{T2_h_mu0} shows the predicted
trends at 3.6\,$\micron$ at $z=0$ for different values of $\lambda$,
representative of those found to fit the multi-wavelength profiles of
nearby disks (\citealt{Munoz-Mateos:2011}). By construction, the
models do not implement the formation of breaks, nor do they allow for
radial transfer of mass. They provide an excellent fit to the
distribution of single-exponential disks in this diagram, and are thus
a useful reference against which to compare down-bending profiles.

As for item 5 in the list above, up to now the connection between
galaxy mass and disk scale-length had been traditionally explored by
assigning a single exponential slope to each galaxy, even though most
disks actually exhibit two distinct slopes. Interestingly,
Fig.~\ref{T2_h_mu0}a demonstrates that this trend still holds true
when the inner and outer exponentials are considered on their own. In
other words, in those galaxies with down-bending profiles, both the
inner and outer disks become shallower in more massive
galaxies. Moreover, even though there is some degree of overlap,
single exponential profiles define a clear boundary in this plot. The
inner slope of a Type~II disk will preferentially be flatter than the
slope of a Type~I disk with the same total mass; similarly, the outer
slope will tend to be steeper. We note that this is not a trivial
result, as there are many $h_i-h_o$ configurations that would yield
the same total mass of a disk with a single scale-length $h$.

Besides the influence of $\lambda$ and of secular processes mentioned
above, this plot also explains part of the scatter in the $h$-$M$
trend found in previous studies. Indeed, when fitting a down-bending
profile with a single exponential function, the resulting scale-length
will depend on the radial position of the break and on the depth of
the image. If the break happens at large radii, then most of the disk
profile will be dominated by the inner disk, so that the fitted
scale-length will be biased towards $h_i$. Conversely, in those disks
where the break occurs closer to the center, the light profile will be
dominated by the outer exponential, thus biasing $h$ towards $h_o$.

\subsubsection{Extrapolated central surface brightness}
Figure~\ref{T2_h_mu0}b shows the extrapolated central surface
brightness, corrected for inclination as
$\mu_\mathrm{corr}=\mu_\mathrm{obs}-2.5 \log (b/a)$. As happened
before with the scale-length, Type~I disks again delineate an obvious
boundary between the inner and outer parts of down-bending
profiles. For a given total stellar mass, the extrapolated central
surface brightness of inner (outer) disks is always fainter (brighter)
than in a similarly massive Type~I disk.

If we focus on single-exponential profiles alone, we can see that for
galaxies brighter than $\sim -20$ (with stellar masses above
$10^{10}\,M_{\sun}$), $\mu_0$ remains rather constant, oscillating
around $\sim 20$\,\maggie, whereas fainter galaxies exhibit dimmer
central surface brightnesses. This constancy of the central surface
brightness of disks was already noted by \cite{Freeman:1970} (the now
called Freeman law). The scatter in that pioneering study was most
likely artificially low due to selection effects, which biased
observations towards high surface brightness galaxies. However, recent
studies carried out on much larger samples of galaxies with deeper
images have corroborated that, albeit with large scatter, $\mu_0$ is
essentially independent of galaxy mass or Hubble type, except for the
latest types. For instance, in a study of the disk structural
properties of roughly $\sim 30,000$ SDSS galaxies, \cite{Fathi:2010}
found that $\mu_0 \sim 20$\,\maggie\ in the $r$ band for early- and
intermediate-type disks; this is fully consistent with our value at
3.6\,$\micron$, considering that $(r-3.6\micron)_{\mathrm{AB}} \sim
0.1-0.2$ for nearby spirals
(\citealt{Munoz-Mateos:2009a}). \cite{Fathi:2010} also noted that
$\mu_0$ drops in galaxies with Hubble types $T \ge 6$ (Scd or
later). We do observe the same trend in Fig.~\ref{T2_h_mu0}, where
Type~I galaxies fainter than $\sim -20$ (mostly late-type ones,
according to Fig.~\ref{sample}) present indeed fainter values of
$\mu_0$.

The Freeman law constitutes yet another constraint for disk evolution
models. In particular, the models of \cite{Boissier:2000} successfully
reproduce the behavior of single-exponential disks (green curves in
Fig.~\ref{T2_h_mu0}b). The central surface brightness increases just
mildly with the total luminosity, as observed. According to the
models, most of the scatter arises from variations in $\lambda$, with
high-spin disks exhibiting fainter surface brightness.

Figure~\ref{T2_h_mu0}b shows that the Freeman law roughly applies not
only to single-exponential disks, but also to the inner component of
down-bending disks. The trend is globally shifted towards fainter
values by 1-2\,mags, but the overall shape is the same: $\mu_{0,i}$
holds roughly constant at $\sim 21$\,\maggie\ for
$M_{\mathrm{3.6\,\micron}} < -20$ and then drops for fainter
disks. This behavior does not seem to be fully mirrored in the outer
disks, though, perhaps because the scatter in $\mu_{0,o}$ at fixed
total stellar mass is significantly larger, around 2-4\,mags. Note
that the extrapolated values of $\mu_{0,o}$ become increasingly more
uncertain in galaxies where the break radius happens at large
galactocentric distances. In these cases, a very small error in the
slope of the outer disk can translate into large variations in the
extrapolated central surface brightness. This can explain part of the
significantly large scatter towards bright values of $\mu_{0,o}$. In
general, the difference between the inner and outer central surface
brightness is $\mu_{0,i} - \mu_{0,o} \sim 1 - 6$\,mags, in very good
agreement with the simulations of \cite{Debattista:2006}.

\subsubsection{Disk isophotal radius}
\begin{figure}
\begin{center}
\resizebox{1\hsize}{!}{\includegraphics{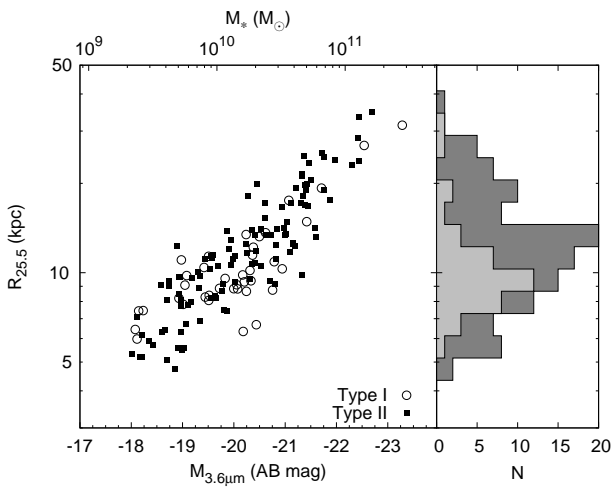}}
\caption{Radius where $\mu_{3.6\,\micron} = 25.5$\,\maggie\ as a
  function of the 3.6\,$\micron$ absolute magnitude, for both
  single-exponential disks (open circles, light histogram) and
  down-bending ones (black squares, dark
  histogram). \label{T2_Rdisk_Mirac1}}
\end{center}
\end{figure}

Besides using disk scale-lengths, another approach to look at galactic
``sizes'' is through the physical radius at a given surface brightness
level. In Fig.~\ref{T2_Rdisk_Mirac1} we plot the radius at our
fiducial level of 25.5\,\maggie\ as a function of absolute
3.6\,$\micron$ magnitude. It can be clearly seen that both
single-exponential and down-bending profiles lie along the same
sequence. This means that if we plot the radial profiles of Type~I and
II disks with the same total stellar mass, they will roughly intersect
at $\mu \sim 25.5$\,\maggie\ if the radius is in physical units.

\subsubsection{Break radius}
\begin{figure}
\begin{center}
\resizebox{1\hsize}{!}{\includegraphics{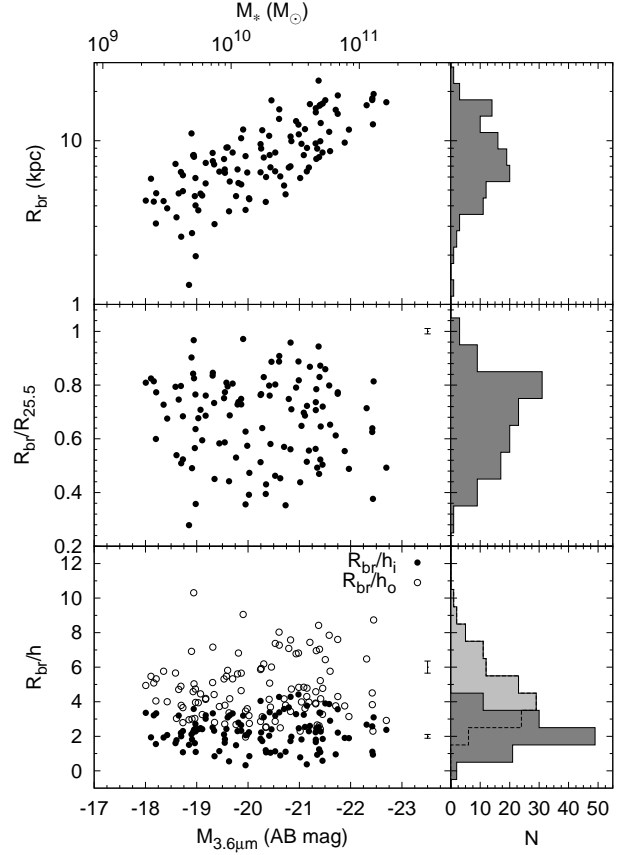}}
\caption{Distribution of the break radius as a function of the
  absolute 3.6\,$\micron$ magnitude. The break radius is shown in
  physical units (top panel), normalized by the radius where
  $\mu_{3.6\,\micron} = 25.5$\,\maggie\ (middle panel), and normalized
  by the inner ($h_i$) and outer ($h_o$) disk scale-length (bottom
  panel). Median error bars are shown in the middle and bottom panels
  (they are  smaller than the symbols in the top one). \label{T2_Rbreak_Mirac1}}
\end{center}
\end{figure}

Where in the disk does the change of slope occur? In
Fig.~\ref{T2_Rbreak_Mirac1} we present the distribution of the break
radius as a function of absolute magnitude. When expressed in kpc,
$R_{\mathrm{br}}$ is clearly correlated with galaxy luminosity,
ranging from roughly 3-4\,kpc for faint galaxies with $M_* \sim 2
\times 10^9 M_{\sun}$ to above 10\,kpc for those with $M_* \sim
10^{11} M_{\sun}$.

When normalizing the break radius by the isophotal diameter
$R_{25.5}$, the trend with mass vanishes completely. The distribution
of data-points mildly peaks at $R\sub{br} \sim 0.8 R_{25.5}$, although
the histogram clearly exhibits extended wings, especially towards
smaller break radii. The asymmetric shape of this histogram, which
drops more abruptly above $R\sub{br} \sim 0.8 R_{25.5}$, can be at
least partly attributed to selection effects: it is obviously harder
to detect breaks at large radii and low surface brightness levels, and
we tend to be more conservative when identifying those breaks.

Finally, the distribution of break radii when normalized to disk
scale-length (both the inner and the outer one) is shown in the bottom
panel of Fig.~\ref{T2_Rbreak_Mirac1}. Again, any trend with mass seems
to be washed out by the scatter. When normalizing to the inner disk
scale-length, the distribution appears to be strongly peaked, with a
mean and rms values of $R\sub{br} = (2.3 \pm 0.9) \times h_i$. The
dispersion is larger in the case of the outer disk, with most breaks
clustered around $R\sub{br} = (4.7 \pm 1.7) \times h_o$. Our empirical
distributions of $R\sub{br}/h_i$ and $R\sub{br}/h_o$ are consistent
with those resulting from the simulations of \cite{Foyle:2008}.

\subsubsection{Stellar surface density at the break}
\begin{figure}
\begin{center}
\resizebox{1\hsize}{!}{\includegraphics{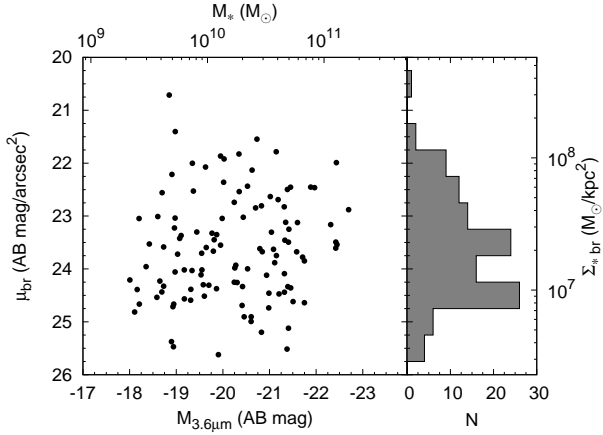}}
\caption{Deprojected 3.6\,$\micron$ surface brightness at the break
  radius as a function of the absolute magnitude of each galaxy. The
  corresponding stellar mass surface density at the break and total
  stellar mass of the galaxy are also
  indicated.\label{T2_mubreak_Mirac1}}
\end{center}
\end{figure}

Do breaks occur at a well defined stellar mass surface density?
Figure~\ref{T2_mubreak_Mirac1} shows the deprojected 3.6\,$\micron$
surface brightness at the position of the break. The distribution is
markedly broad, with no evident dependence on the total stellar mass
of the galaxy. In general, most breaks can be found anywhere in the
range $\mu_{\mathrm{br}} \sim 22 - 25$\,\maggie\ or, equivalently,
$\Sigma_{*} \sim 5 \times 10^7 - 10^8\,M_{\sun}\,\mathrm{kpc}^{-2}
$. In their simulations of break formation and evolution,
\cite{Foyle:2008} found that the total baryonic (gas + stars) surface
density at the break radius typically lied between $10^7 -
10^8\,M_{\sun}\,\mathrm{kpc}^{-2}$, whereas the gas surface density
alone ranged between $10^6 -
10^7\,M_{\sun}\,\mathrm{kpc}^{-2}$. Subtracting the latter from the
former yields a distribution of stellar mass surface density at the
break radius entirely consistent with our findings.

\subsection{Global properties of bars}
Now that we have broadly described the main structural properties of
down-bending profiles, and compared them with equivalent
single-exponential ones with the same stellar mass, we will now
proceed to investigate the potential role of bars in shaping these
disks. We begin with an overview of the structural properties of bars
in our sample.

\begin{figure}
\begin{center}
\resizebox{1\hsize}{!}{\includegraphics{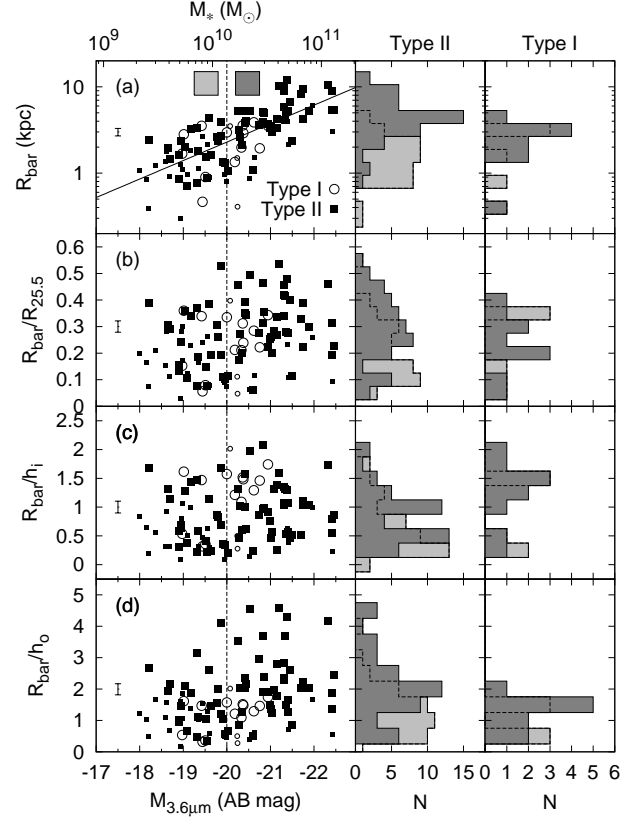}}
\caption{Bar radius as a function of absolute 3.6\,$\micron$
  magnitude. From top to bottom, the bar radius is shown (a): in kpc,
  (b): normalized by the radius where $\mu_{3.6\,\micron} =
  25.5$\,\maggie, (c): normalized by the scale-length of the inner
  disk ($h_i$), and (d): by that of the outer disk
  ($h_o$). Down-bending profiles are marked with solid squares,
  whereas single-exponential ones are shown with open circles. Small
  symbols are used in either case to identify ``candidate'' bars. The
  histograms to the right show the distribution of bar radii after
  having divided our sample in two bins: galaxies fainter than $-20$
  (light gray) and brighter (dark gray). Candidate bars are included
  in these histograms. Note that the values of $R_{\mathrm{bar}}/h_i$
  and $R_{\mathrm{bar}}/h_o$ are exactly the same for Type~I profiles,
  since they have a single scale-length. Median error bars are
  shown.\label{T2_Rbar_Mirac1}}
\end{center}
\end{figure}

In Fig.~\ref{T2_Rbar_Mirac1} we show the distribution of bar radii as
a function of galaxy luminosity. For the sake of comparison, we plot
the bar radii of both single-exponential (open circles) and
down-bending profiles (solid squares). In order to distinguish
``candidate'' bars from ``genuine'' ones, we tag the former with
smaller symbols. We have also divided our sample in two bins of bright
and faint galaxies, taking $M_{\mathrm{3.6\,\micron}} = -20$ ($M_*
\sim 10^{10} M_{\sun}$) as a limiting boundary, shown in the plot with a
vertical dashed line. The histograms to the right describe the
distribution of bar radii for the bright and faint galaxies using a
dark and light shade of gray, respectively.

Our trends between bar size and galactic mass are fully consistent
with previous optical and near-IR studies of bars in the local
universe (\citealt{Martin:1995, Laurikainen:2002, Erwin:2005a,
  Menendez-Delmestre:2007}). In panel (a) we can clearly see that more
massive galaxies also host longer bars. This is not surprising,
though, since bar length is known to correlate with disk scale-length
or size (e.g.$.$ \citealt{Elmegreen:1985}) and larger disks tend to be
more massive. The linear fit in this panel corresponds to:
\begin{equation}
\log R\sub{bar}\ (\mathrm{kpc})=-3.920-0.214
M_{\mathrm{3.6\,\micron}}\ (\mathrm{AB})\label{eq_rbar}
\end{equation}
with a $1\sigma$ scatter of $\sim 0.23$\,dex. We will make use of this
relation later in Section~\ref{resonances} when predicting the locus
of different resonances.

Panel (b) shows the distribution of bar radii normalized by our
reference $R_{25.5}$ radius at 3.6\,$\micron$. Despite the scatter, a
trend with mass is still visible, in the sense that bars in massive
disks are longer relative to the overall disk size. The histograms
demonstrate that in massive disks, bars typically have
$R_{\mathrm{bar}} \sim 0.3 R_{25.5}$, while in less massive ones they
tend to be half as long, with $R_{\mathrm{bar}} \sim 0.15 R_{25.5}$
(\citealt{Elmegreen:1985}).

The trends are not so clear, but yet still present, when using disk
scale-length as a measuring rod against which to compare the bar size
(panels c and d). Although with large scatter, bars in massive disks
typically extend out to $0.5-1 \times h_i$ and $2 \times h_o$, whereas
those in low mass disks reach out to $0.25 \times h_i$ and $0.5-1
\times h_o$. 

There is no evident difference in the distribution of bar radii
between Type~I and II disks. There is a hint from panels (a) and (b)
that, if we restrict ourselves to faint disks (light histograms), bars
would be $\sim 3$ times shorter in down-bending profiles than in
single exponential ones. This is very hard to ascertain, though, given
that purely exponential profiles are so rare. Also, Type~I disks lie
in the upper and lower parts of panels (c) and (d), respectively,
because their unique scale-length $h$ is intermediate between $h_i$
and $h_o$ for a given stellar mass, as shown in
Fig.~\ref{T2_h_mu0}a.

\subsection{The connection between bars and breaks}\label{resonances}
\begin{figure}
\begin{center}
\resizebox{1\hsize}{!}{\includegraphics{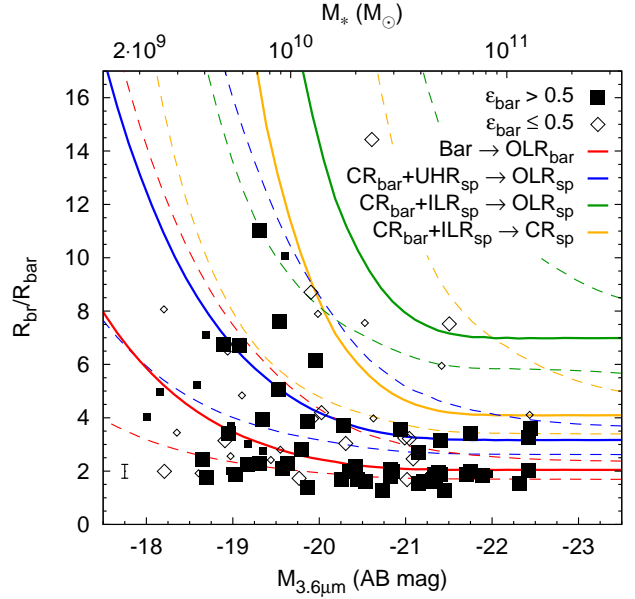}}
\caption{Break radius relative to the bar radius, as a function of
  absolute 3.6\,$\micron$ magnitude. Filled squares and open diamonds
  correspond to galaxies with high- and low-ellipticity bars,
  respectively. In both cases, small symbols denote candidate
  bars. The median error bar is shown in the bottom-left part. The red
  solid curve tracks the expected locus of $R\sub{OLR}/R\sub{bar}$,
  assuming a realistic rotation curve rather than a flat one. The blue
  one corresponds to the case where the inner 4:1 resonance of the
  spiral pattern overlaps with the bar CR, and the break happens at
  the spiral OLR. The green curve shows the case where the bar CR
  coincides with the spiral ILR, with the break being again at the
  spiral OLR. The orange curve has the same coupling as the green one,
  but assuming that the break happens at the spiral CR instead. Dashed
  lines account for uncertainties in the involved variables (see text
  for more details).\label{T2_Rbreak_Rbar}}
\end{center}
\end{figure}

Does the bar determine the radius of the break? In
Fig.~\ref{T2_Rbreak_Rbar} we plot the
$R_{\mathrm{br}}/R_{\mathrm{bar}}$ ratio as a function of galaxy
luminosity. This diagram demonstrates that the range of possible
break-to-bar ratios is strongly dependent on the total stellar
mass. Galaxies fainter than $-20$ (with stellar masses below $10^{10}
M_{\sun}$) exhibit breaks at galactocentric distances ranging from 2
to more than 10 times the bar radius. More massive disks, on the other
hand, are clustered around $R_{\mathrm{br}}/R_{\mathrm{bar}} \sim
2-3$; the very few ones with larger break radii tend to host bars with
ellipticities lower than 0.5, and are thus presumably weak bars.

From a purely observational perspective, this is consistent with the
results of previous work (\citealt{Pohlen:2006}; E08), where ``OLR
breaks'' (those with $R_{\mathrm{br}}/R_{\mathrm{bar}} \sim 2-3$) were
found to be more common in early-type disks, while ``classical
breaks'' (at larger radii) were more abundant in late-type
disks. However, we believe that this dichotomy might be too simple to
fully encompass the wealth of features shown in
Fig.~\ref{T2_Rbreak_Rbar}. Moreover, by presuming a distinct physical
origin for ``OLR'' and ``classical'' breaks (resonances versus SF
thresholds), this nomenclature could be clouding our understanding of
the actual physics behind these features.

\begin{figure}
\begin{center}
\resizebox{1\hsize}{!}{\includegraphics{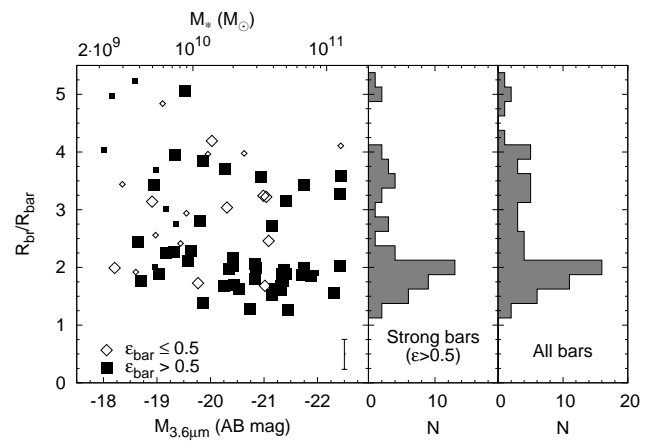}}
\caption{Enlarged plot of the lower region in
  Fig.~\ref{T2_Rbreak_Rbar}. The median error in
  $R_{\mathrm{br}}/R_{\mathrm{bar}}$ is shown at the bottom-right
  part of the plot.\label{T2_Rbreak_Rbar_zoom}}
\end{center}
\end{figure}

Two particular issues should be noted in this regard:
\begin{enumerate}
\item Besides the classical OLR of the bar alone, other resonances
  might be involved in creating breaks. There is tantalizing evidence
  for this in Fig.~\ref{T2_Rbreak_Rbar_zoom}, which highlights the
  lower region of Fig.~\ref{T2_Rbreak_Rbar}. This plot shows that the
  distribution of $R_{\mathrm{br}}/R_{\mathrm{bar}}$ seems to be
  bimodal, with two separate sequences of data-points clustering
  around {\it either} $R_{\mathrm{br}}/R_{\mathrm{bar}} \sim 2$ or
  $\sim 3.5$. This might be a tell-tale sign that more than one set of
  resonances is at play.

\item The fact that in some cases $R_{\mathrm{br}}/R_{\mathrm{bar}}
  \gg 2-3$, especially in low-mass disks, does not necessarily imply
  that breaks arise from a SF-related mechanism rather than from a
  dynamical one. It is perfectly possible for a resonance to be found
  at large radii in these systems, as we will show.
\end{enumerate}

\subsubsection{Bar-only resonances}
To address these issues in more detail, let us consider a disk galaxy
with a given rotation curve $V(r)=r \Omega (r)$. To first order, the
orbit of a star can be described as the superposition of a circular
orbit, with an angular velocity $\Omega (r)$, and a smaller elliptical
epicycle around it, with an angular frequency $\kappa (r)$. This
epicyclic frequency is given by (see, e.g., \citealt{Binney:2008} and
references therein):
\begin{equation}
\kappa^2 = 2 \Omega \left[2 \Omega +r (d \Omega/dr)\right]
\end{equation}

When the disk hosts a non-axisymmetric pattern, such as a bar or a
spiral structure, that rotates as a solid body with a given pattern
speed $\Omega_p$, several important resonances can be found. The
corotation resonance (CR) occurs at the radius where stars rotate with
the same angular velocity as the perturbing pattern, $\Omega =
\Omega_p$. The Lindblad Resonances are found where a star completes
one epicycle between consecutive encounters with the pattern; that is,
where $\Omega_p = \Omega \pm k/m$, with $m$ being the multiplicity of
the pattern (2 for a bar or a two-armed spiral). The positive sign
denotes the Outer Lindblad Resonance (OLR), whereas the negative one
corresponds to the Inner Lindblad Resonance (ILR).

Under the assumption of a flat rotation curve, it is clear from the
definition above that $\kappa=\sqrt{2}V/r$, and therefore
$R\sub{OLR}/R\sub{CR}=1+1/\sqrt{2} \simeq 1.7$. To relate this to the
bar radius, we need to assume a particular value for $\RR \equiv
R\sub{CR}/R\sub{bar}$. Several techniques have been developed over the
years to measure this ratio. These include, among others, the
Tremaine-Weinberg method (\citealt{Tremaine:1984}), matching gas flow
models with observations (\citealt{Sanders:1980, Lindblad:1996,
  Weiner:2001}) and identifying resonances with certain morphological
features such as rings or dust lanes (\citealt{Buta:1986,
  Athanassoula:1992, Elmegreen:1992, Moore:1995, Perez:2012}). Despite
building on completely different methodologies and assumptions, these
techniques generally yield consistent results. They normally point
towards $\RR \simeq 1.2 \pm 0.2$, with the bar ending inside but close
to corotation. This implies that $R\sub{OLR}/R\sub{bar} \simeq 2$,
which can nicely explain the large number of breaks found at twice the
bar radius, as already noted in previous works.

However, it is worth emphasizing that this estimate relies heavily on
the assumption of a flat rotation curve, which is reasonable for
massive disks, but might be quite far-fetched for low-mass ones, where
the rotation curve rises gently for a large fraction of the optical
disk. Indeed, \cite{Elmegreen:1985} showed that bars in early-type
disks extend well beyond the rising part of the rotation curve,
whereas bars in late-type disks end before the velocity flattens.

The HI Nearby Galaxies Survey (THINGS; \citealt{Walter:2008}) has
provided HI maps of unprecedented quality for a representative set of
nearby galaxies. Rotation curves for these objects were presented by
\cite{de-Blok:2008}, and later fitted by \cite{Leroy:2008} using the
following analytical expression:
\begin{equation}
V(r)=V\sub{flat} (1-e^{-r/r\sub{flat}})
\end{equation}
where $V\sub{flat}$ is the asymptotic rotation velocity, and $r\sub{flat}$
is the radial scale over which the flat regime is
reached\footnote{See, e.g., \cite{Athanassoula:1982} for a similar
  derivation of resonance radii with a different mathematical
  parameterization of the rotation curve.}. We fitted $r\sub{flat}$ as a
function of the absolute magnitude at 3.6\,$\micron$, taken from
\cite{Munoz-Mateos:2009a}. In general, $r\sub{flat}$ is
around 1\,kpc, and increases mildly with decreasing mass. Using this
information, we can build more realistic curves of $\Omega(r)$ and
$\kappa(r)$ tailored for each absolute magnitude.

From the fit in the top panel of Fig.~\ref{T2_Rbar_Mirac1}
(Eq.~\ref{eq_rbar}), we can get the typical bar radius for a galaxy of
a given absolute magnitude. Multiplying this value by $\RR = 1.2$ we
obtain an estimate of the CR radius, which in turn yields the
corresponding OLR radius when combined with the proper rotation curve
for that particular absolute magnitude, as explained above. The
resulting prediction of $R\sub{OLR}/R\sub{bar}$ is shown as a red line
in Fig.~\ref{T2_Rbreak_Rbar}. As expected, in the limit of high-mass
galaxies we recover the value of $\sim 2$ that we derived before,
since in these galaxies $R\sub{bar} \gg r\sub{flat}$. However, the OLR
is found further away from the bar in low mass disks, where the rising
nature of the rotation curve cannot be ignored given that $R\sub{bar}
\lesssim r\sub{flat}$.

The dashed curves mark the estimated uncertainty in this prediction,
resulting from the $1\sigma$ scatter in the involved variables,
namely:
\begin{enumerate}
\item Scatter in $R\sub{bar}$ for a given $M_{\mathrm{3.6\,\micron}}$:
  $\pm 0.23$\,dex (Fig.~\ref{T2_Rbar_Mirac1}).
\item Scatter in $r\sub{flat}$ for a given $M_{\mathrm{3.6\,\micron}}$:
  $\pm 0.4$\,dex (from the data in \citealt{Leroy:2008})
\item Scatter in $\RR$: $\pm 0.2$
  (\citealt{Athanassoula:1992, Elmegreen:1996, Aguerri:2003, Corsini:2011}).
\end{enumerate}
For simplicity, we have assumed that there is no degree of correlation
or anti-correlation in the scatter among these variables. In practice,
though, the dashed curves could be somewhat different if this
assumption is not valid.

This physically motivated model shows that at least part of the
scatter in $R\sub{br}/R\sub{bar}$ could be simply due to the different
dynamical properties of low-mass disks. Breaks in these galaxies could
be perfectly well linked to the OLR of the bar; this resonance just
happens to be placed further away from the bar due to the rising
rotation curve of these objects.

An additional factor that we have not considered here is the fact that
bars in some low-mass disks could be slow rotators, with $\RR > 1.4$
(\citealt{Rautiainen:2005, Meidt:2009} and references therein). Large
values of $\RR$ in these objects would further increase the OLR-to-bar
ratio, more than what is already depicted by the red curves.

\subsubsection{Coupled spiral-bar resonances}
As we mentioned before, an intriguing feature of
Figs.~\ref{T2_Rbreak_Rbar} and \ref{T2_Rbreak_Rbar_zoom} is what
appears to be a second family of galaxies having breaks at roughly
$3.5 R\sub{bar}$. One possible scenario worth exploring is the
possibility that this reflects a coupling between the bar and spiral
patterns (\citealt{Tagger:1987, Masset:1997, Sygnet:1988,
  Rautiainen:1999, Quillen:2011, Minchev:2011}). Such a coupling might
exist if some resonances of the bar and spiral structure overlap. In
this case, the radial transfer of angular momentum can proceed further
out, leading to breaks at some point inside the OLR of the spiral
rather than the OLR of the bar.

We can consider the case in which the inner 4:1 resonance of the
spiral pattern (also known as the Ultra Harmonic Resonance, UHR)
overlaps with the bar corotation. Under the assumption of a flat
rotation curve, one can see that the OLR of the spiral and the CR
radius of the bar are such that
$R\sub{OLR,sp}/R\sub{CR,bar}=(1+1/\sqrt{2})/(1-\sqrt{2}/4) \simeq
2.6$. Again, if we suppose that $\RR \simeq 1.2$, then this yields
$R\sub{OLR,sp}/R\sub{bar} \simeq 3.1$, consistent with the value of
$R\sub{br}/R\sub{bar}$ that we observe.

As we did before in the case of a bar-only OLR, we can extend this
result to the generic case of a realistic rotation curve, where the
radial scale $r\sub{flat}$ over which the curve is rising depends on
the absolute magnitude of the galaxy. The resulting predicted locus
for $R\sub{OLR,sp}/R\sub{bar}$ is marked with a blue curve in
Fig.~\ref{T2_Rbreak_Rbar}. Again, the dashed lines show the impact of
the observed 1$\sigma$ variations in $R\sub{bar}$, $r\sub{flat}$ and
$\RR$. This figure demonstrates that, at least in principle,
resonances can account for most of the observed variations in
$R\sub{br}/R\sub{bar}$ as a function of mass, without appealing to a
different mechanism for the formation of the break, such as a SF
threshold. This does not necessarily mean that such thresholds do not
play a role in forming breaks, but warns against systematically
dismissing resonances by default whenever the break happens at more
than $\sim 2 R_{\mathrm{bar}}$.

The particular coupling between the bar and the spiral pattern
described above is just one of several possible scenarios. For
instance, if the bar CR overlaps with the spiral ILR, then the spiral
OLR would be located even further out, as shown by the green curves in
Fig.~\ref{T2_Rbreak_Rbar}. With this coupling, breaks would form at
$\simeq 7 R\sub{bar}$ for a flat rotation curve, and further out for a
rising one. Objects in this area of the plot have low-ellipticity bars
(some are even just candidate bars), so this scenario might not
necessarily apply to them, but it is still a possibility worth
considering.

Some simulations show that under a bar-spiral coupling the break can
form at the spiral CR rather than at the OLR
(\citealt{Minchev:2012}). The orange curve illustrates this, by
assuming the same coupling as the green one, but placing the break at
the spiral CR. The limiting value for a flat rotation curve
($R\sub{br} \simeq 4 R\sub{bar}$) somewhat overpredicts the break
radius for most of our massive disks, but several of our
intermediate-mass disks with $M_{*} \simeq 10^{10} M_{\sun}$ and
$R\sub{br} \simeq 8 R\sub{bar}$ could be consistent with this
scenario. In brief, it is worth emphasizing that given the multiple
ways in which the resonances of bars and spirals can overlap, the
resulting distribution of break radii can be more complex than
previously assumed.

The resonant coupling between the bar and the spiral pattern discussed
above has important consequences for the redistribution of angular
momentum in galactic disks. In some of their numerical simulations of
disks, \cite{Debattista:2006} noted that the bar and the spiral were
indeed coupled via an overlap of the bar CR and the spiral UHR, like
the one we propose here. In those cases, they found that the disk
break took place inside but close to the OLR of the spiral, which is
consistent with our results (blue curve in
Fig.~\ref{T2_Rbreak_Rbar}). \cite{Minchev:2010} highlighted the impact
of spiral-bar coupling in the context of radial stellar
migration. They noted that the effect of such coupling in the
redistribution of angular momentum is highly non-linear, in the sense
that increasing the amplitude of the bar and spiral perturbers has a
significantly larger impact than considering each perturber
separately. As a result, this mechanism seems to be quite efficient in
driving radial stellar migration. For simulations tuned to mimic the
properties of a MW-like galaxy, \cite{Minchev:2010} found that after
just $\sim 3$\,Gyr the bar-spiral coupling yields the same degree of
mixing for which other mechanisms such as transient spirals would take
three times longer (see below). In a follow-up study,
\cite{Minchev:2012} showed that this resonant coupling can in fact
produce breaks at large galactocentric radii, without invoking star
formation thresholds.

Studies on radial migration triggered by coupling between different
patterns usually concentrate on steady state spirals, that is, spirals
that are stable over at least several rotations so that they can
dynamically couple with other patterns such as the bar. Transient
spiral arms constitute an alternative mechanism to drive stellar
migration, as proposed by \cite{Sellwood:2002} (see also
\citealt{Roskar:2012}). Under this scenario, stars are sent from the
CR of one spiral pattern to another, and the transient nature of the
spirals prevents stars from being trapped at certain
orbits. Nevertheless, these two mechanisms (transient spirals and
bar-spiral coupling) are not necessarily exclusive; in fact, as
discussed by \cite{Minchev:2010} and \cite{Minchev:2012}, due to its
non-linear nature, pattern coupling works with both long- and
short-lived spirals.

From an observational point of view, resonant coupling between
different patterns might be detected in actual galaxies using the
Radial Tremaine-Weinberg method (\citealt{Merrifield:2006}), a
generalized modification of the TW method that allows for radial
variations in the pattern speed. \cite{Meidt:2009} applied this
technique to a sample of nearby galaxies, and found indeed suggestive
evidence for resonant coupling in some of them. Given the tantalizing
signatures of potential pattern coupling seen in
Fig.~\ref{T2_Rbreak_Rbar}, this kind of more detailed analysis is
definitely worth pursuing in future papers.

Can this scenario be applied to unbarred galaxies as well? Spiral
structure and spiral-spiral coupling can take place in the absence of
a bar (\citealt{Sygnet:1988, Rautiainen:1999}), so one should also
consider spiral resonances as a possible mechanism for the formation
of breaks in unbarred galaxies, besides SF thresholds. Such an
analysis is beyond the scope of the current paper, but will be
addressed in future work.

\subsection{What triggers the onset of a break?}
\cite{Foyle:2008} proposed that whether a galaxy develops a break
or not is largely determined by the ratio $m\sub{d}/\lambda$, where
$m\sub{d}$ is the disk-to-halo mass fraction and $\lambda$ is the
dimensionless spin parameter. In their simulations, galaxies with
$m\sub{d}/\lambda \geq 1$ ended up forming a break, whereas those with
a low $m\sub{d}/\lambda$ never developed one.

Confronting this hypothesis against observations is not
straightforward. On one hand, computing $m\sub{d}$ requires accurate
rotation curves, which are only available for a subset of our
galaxies. On the other hand, $\lambda$ cannot be directly measured
from observations. While it can be {\it indirectly} inferred from the
light profiles of galaxies (\citealt{Munoz-Mateos:2011} and references
therein), this requires neglecting radial stellar migration, which we
now know can significantly reshape the light profiles of galaxies and
bias the inferred values of $\lambda$.

Nevertheless, \cite{Foyle:2008} noted that the concentration index of
their simulated galaxies was a reasonable indicator of wheher a disk
would develop a break. For each galaxy they measured the temporal
evolution of $C_{28} = 5 \log (r_{80}/r_{20})$, where $r_{80}$ and
$r_{20}$ are the radii encompassing 20\% and 80\% of the total
baryonic mass (\citealt{Kent:1985}). They found that after 5\,Gyrs
almost all galaxies with breaks had $C_{28} > 4$, whereas those with
lower concentrations remained as single-exponential disks. After
10\,Gyrs the concentration index separating broken and unbroken
profiles had dropped to $C_{28} \sim 3.5$.

\begin{figure*}
\begin{center}
\resizebox{1\hsize}{!}{\includegraphics{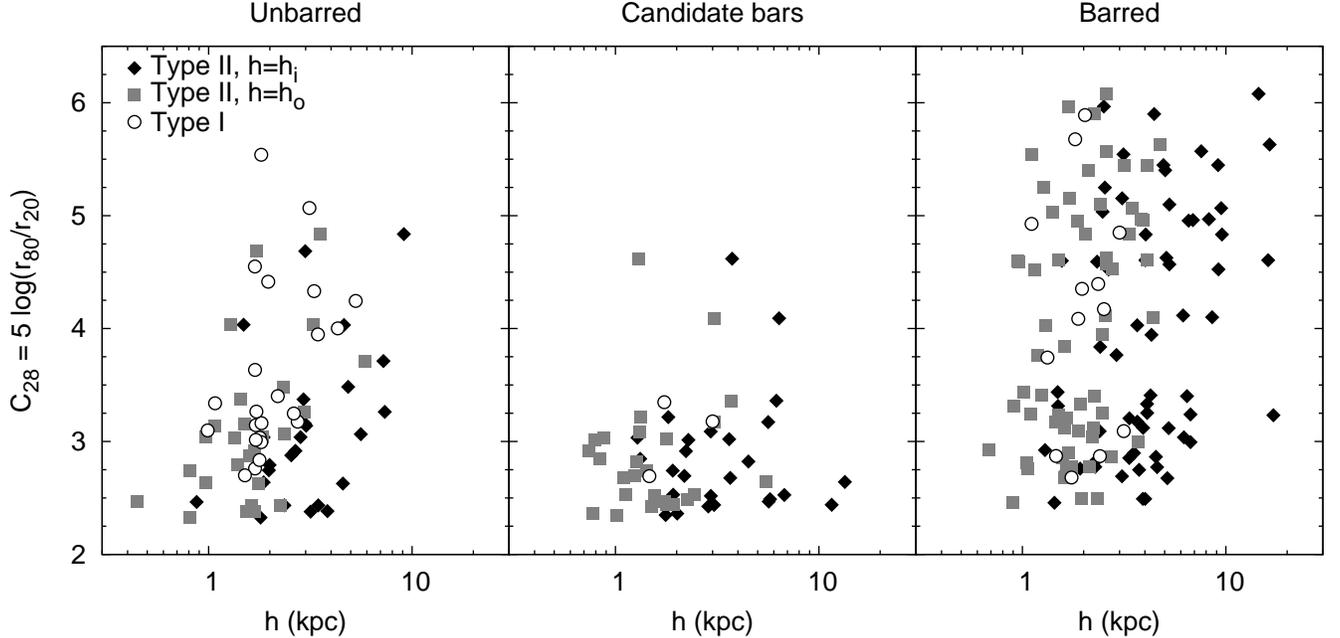}}
\caption{Concentration index of the 3.6\,$\micron$ profiles as a
  function of the disk scale-length. Single-exponential profiles are
  shown with open circles. Down-bending profiles are represented by
  two data-points each, using both the inner disk scale-length as the
  abscissa (black diamonds) or the outer one (gray squares). Galaxies
  have been sorted out into unbarred, candidate bars and
  barred.\label{T2_cindex}}
\end{center}
\end{figure*}

In Fig.~\ref{T2_cindex} we plot $C_{28}$ (as measured on the
3.6\,$\micron$ profiles) as a function of the disk scale-length for
broken profiles (filled symbols) and unbroken ones (open symbols). We
also consider separately barred galaxies, unbarred ones and candidate
bars. A perfect exponential profile has $C_{28}=2.7$, but the presence
of a central bulge and/or bar increases this value, as can be readily
seen here. This plot should be compared with Fig.~12 in
\cite{Foyle:2008}\footnote{Note that these authors used the {\it
    initial} disk scale-length of their simulated disks as their
  x-axis, which we obviously cannot do with our actual galaxies. This
  is not extremely relevant, though, as the parameter of interest here
  is $C_{28}$ along the y-axis. \cite{Foyle:2008} measured this
  concentration index at 5 and 10\,Gyrs on the total baryonic profiles
  (gas + all stars). Throughout their simulations the gas component
  represents a small contribution, so our stellar values of $C_{28}$
  at 3.6\,$\micron$ constitute a good proxy.}. In contrast with the
results of that study, here we do not find that $C_{28}$ segregates
broken and unbroken disks. In general, we find disks with and without
breaks both above and below the theoretical limit of $C_{28} \simeq
3.5-4$ resulting from their simulations. If $C_{28}$ maps
$m\sub{d}/\lambda$ (at least to first order), then our results do not
support the idea that such ratio governs the onset of disk breaks.

\section{Conclusions}\label{conclusions}
In this work we have carried out a detailed study of the radial
distribution of old stars in 218 nearby almost face-on disks, using
deep 3.6\,$\micron$ images from \sfg. In particular, we have
investigated the structural properties of disk breaks and their
connection with bars, in order to verify whether non-axisymmetric
perturbations can lead to the observed breaks through secular
rearrangement of stars and angular momentum. Our results can be
summarized as follows:

\begin{itemize}

\item The scale-length of both the inner disk ($h_i$) and the outer
  one ($h_o$) increases monotonically with increasing total stellar
  mass. Interestingly, galaxies with a genuine single-exponential disk
  have a scale-length which lies between those of the inner and outer
  disk of down-bending profiles with the same total stellar mass. On
  average we find $h_i/h_o \simeq 1 - 3$, which agrees with the grid
  of simulations by \cite{Debattista:2006} and \cite{Foyle:2008},
  among others.

\item The extrapolated central surface brightness of the inner
  ($\mu_{0,i}$) and outer disks ($\mu_{0,o}$) do not exhibit any clear
  trend with the total stellar mass. In this regard, they mimic the
  Freeman law for single-exponential disks. Again, for a given stellar
  mass, the central surface brightness of a single-exponential disk is
  intermediate between those of the inner and outer components of a
  down-bending disk. We typically measure $\mu_{0,i} - \mu_{0,o} \sim
  1 - 6$\,mags, in agreement with the simulations of, e.g.,
  \cite{Debattista:2006}.

\item The break radius ranges from 3-4\,kpc for galaxies with $M_*
  \sim 2 \times 10^9 M_{\sun}$ to 10-20\,kpc in those with $M_* \sim
  10^{11} M_{\sun}$. However, the trend vanishes when normalizing the
  break radius by the isophotal size of the disk or by the disk
  scale-length. On average, most breaks occur at $R_{\mathrm{br}} =
  (2.3 \pm 0.9) \times h_i$ and $(4.7 \pm 1.7) \times h_o$. The
  stellar mass surface density at the break radius ranges between
  $\sim 5 \times 10^7 - 10^8\,M_{\sun}\,\mathrm{kpc}^{-2} $, with no
  trend with the mass of the host galaxy. Both the break radius and
  its stellar mass density agree with values found in numerical
  simulations (\citealt{Debattista:2006, Foyle:2008, Roskar:2008,
    Sanchez-Blazquez:2009, Minchev:2012}).

\item In the particular case of barred galaxies, the ratio between the
break and the bar radii presents an intriguing dependence on the total
stellar mass of the galaxy. In objects less massive than
$10^{10}\,M_{\sun}$, breaks can be found between 2 and $\sim$10 times
the bar radius. However, the scatter decreases considerably in more
massive disks, where most breaks lie within 4 bar radii. This behavior
is most pronounced when we consider only highly elongated bars.

\item The distribution of $R_{\mathrm{br}}/R_{\mathrm{bar}}$ in
  massive disks seems to be in fact bimodal. In most cases, the break
  takes place at twice the bar radius, which is the expected locus of
  the Outer Lindblad Resonance of the bar, under the assumption of a
  flat rotation curve. However, there appears to be a second family of
  disks with breaks at $\sim 3.5 R_{\mathrm{bar}}$. We have shown that
  if the bar co-rotation radius overlaps with the inner 4:1 resonance
  of the spiral pattern, breaks are indeed expected to form at that
  radius, in agreement with numerical simulations
  (\citealt{Debattista:2006}). Other combinations of resonances can
  produce breaks at even larger radii. Such resonant coupling between
  different patterns is most relevant in the context of secular
  evolution of disks, since numerical simulations have demonstrated
  that radial stellar migration becomes more efficient in this case
  (\citealt{Minchev:2010, Minchev:2011, Minchev:2012}).

\item It is normally assumed that resonances cannot be responsible for
  those breaks found at $R_{\mathrm{br}} \gg 2 \times
  R_{\mathrm{bar}}$, and other mechanisms such as SF thresholds are
  usually invoked. However, this kind of breaks occur mostly in
  low-mass disks, where bars end before the rotation curve has reached
  the flat regime. This tends to push the aforementioned resonances
  further out, compared to the case of a flat rotation curve, which is
  more suitable for massive disks. Using average but realistic
  rotation curves as a function of the total stellar mass, we have
  shown that resonances with the bar and/or the spiral pattern can
  account for the increased scatter in
  $R_{\mathrm{br}}/R_{\mathrm{bar}}$ at low masses. While this does
  not rule out SF thresholds as a break formation mechanism, it
  cautions against discarding resonances by default whenever a break
  happens far away from the bar.

\item It has been proposed that the development of breaks is governed
  by $m\sub{d}/\lambda$, the ratio between the disk mass fraction and
  the halo spin parameter (\citealt{Foyle:2008}). While such quantity
  cannot be easily measured, the light concentration index $C_{28}$
  has been brought forward as a reasonable proxy. We find, however, no
  connection between high/low values of $C_{28}$ and the
  presence/absence of a break.
\end{itemize}

\acknowledgments The authors are grateful to the entire \sfg\ team for
their collective effort in this project. We thank S$.$ Boissier for
allowing us to use his disk evolution models. We also thank the
referee for a very constructive report that helped to improve the
quality of the paper. JCMM acknowledges financial support from NASA
JPL/Spitzer grant RSA 1374189 provided for the \sfg\ project. JCMM, KS
and TK acknowledge support from the National Radio Astronomy
Observatory, which is a facility of the National Science Foundation
operated under cooperative agreement by Associated Universities,
Inc. We acknowledge financial support from the People Programme (Marie
Curie Actions) of the European Union's FP7/2007-2013/ to the DAGAL
network under REA grant agreement number PITN-GA-2011-289313.

This work is based on data acquired with the Spitzer Space Telescope,
and makes extensive use of the NASA/IPAC Extragalactic Database (NED),
both of which are operated by the Jet Propulsion Laboratory,
California Institute of Technology under a contract with the National
Aeronautics and Space Administration (NASA). We also acknowledge the
usage of the HyperLeda database (http://leda.univ-lyon1.fr).

{\it Facilities:} \facility{{\it Spitzer}}

\appendix
\section{Mass-to-light ratio at 3.6\,$\micron$}\label{ML_ratio}
Throughout this paper we have relied on the mass-to-light ratio at
3.6\,$\micron$ of \cite{Eskew:2012}. These authors compared
spatially-resolved stellar mass maps of the Large Magellanic Cloud
with the corresponding IRAC maps to derive the following calibration:
\begin{equation}
\frac{M_*}{M_{\sun}} = 10^{5.97} \frac{F_{3.6}}{\mathrm{Jy}} \left(\frac{D/\mathrm{Mpc}}{0.05}\right)^2
\end{equation}
or, equivalently:
\begin{equation}
\log{M_*/M_{\sun}} = -0.4 M_{3.6\,AB}+2.13\label{eq_ML_1band}
\end{equation}

They also provide the following calibration when fluxes at both
3.6\,$\micron$ and 4.5\,$\micron$ are available:
\begin{equation}
\frac{M_*}{M_{\sun}} = 10^{5.65} \frac{F_{3.6}^{2.85}}{\mathrm{Jy}} \frac{F_{4.5}^{-1.85}}{\mathrm{Jy}}\left(\frac{D/\mathrm{Mpc}}{0.05}\right)^2\label{eq_ML_2bands}
\end{equation}
Galaxies in our sample exhibit a very narrow range of $3.6-4.5$
colors, the average value being $-0.41 \pm 0.08$\,mags
(AB). Substituting this value into Eq.~\ref{eq_ML_2bands} yields:
\begin{equation}
\log{M_*/M_{\sun}} = -0.4 M_{3.6\,AB}+2.12 \pm 0.06
\end{equation}
which is entirely consistent with Eq.~\ref{eq_ML_1band}. The
corresponding stellar mass surface density is given by:
\begin{equation}
\log \Sigma_{*} (M_{\sun}\,\mathrm{kpc}^{-2}) = 16.76-0.4 \mu_{3.6} (\mathrm{AB\,mag\,arcsec}^{-2})
\end{equation}

\section{Radial profiles}\label{atlas}
To facilitate the visual inspection of the structural properties of
our galaxies, in this appendix we compile the images and radial
profiles for the whole sample, overplotting the position of the breaks
and bars when present. The whole set of figures can be found in the
electronic edition of the journal.
\begin{figure}
\begin{center}
\resizebox{1\hsize}{!}{\includegraphics{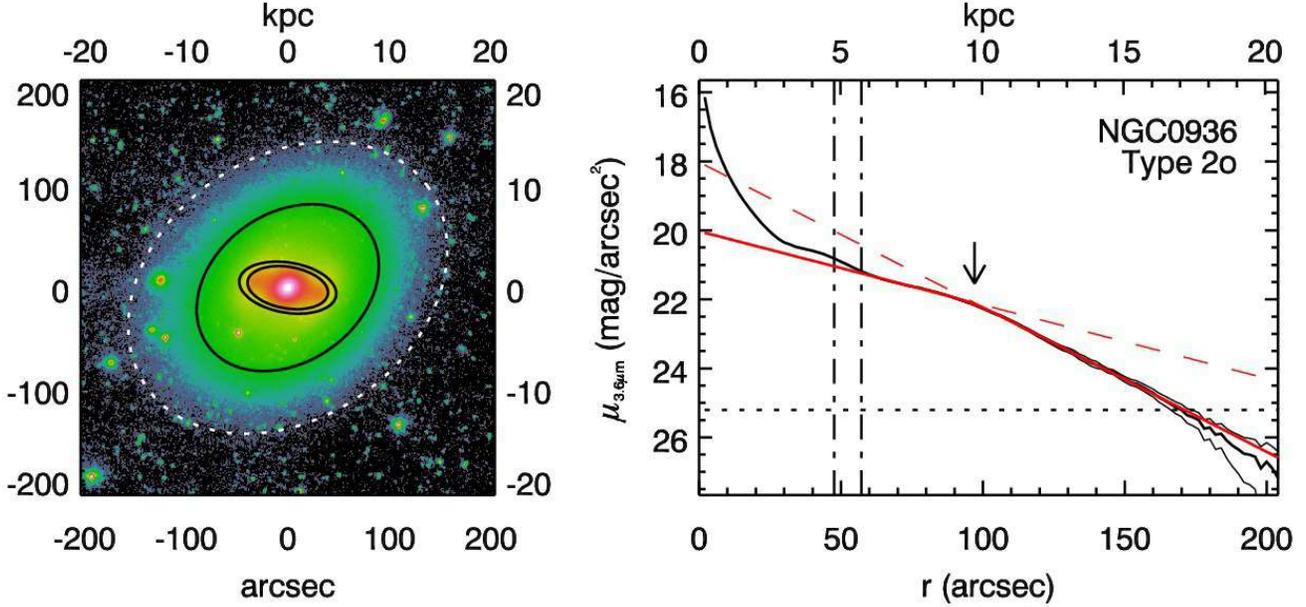}}
\caption{Image and surface brightness profile for NGC~0936 at
  3.6\,$\micron$. The best-fitting disk model
  (Sect.~\ref{disk_measurements}) is shown as a red solid curve; for
  clarity, red dashed lines indicate the extrapolation of the inner
  and outer components of the profile. The position of the break is
  marked with a vertical arrow in the profile and a black solid
  ellipse in the image. The limits for the bar radius
  (Sect.~\ref{bar_measurements}) are shown as two vertical dash-dot
  lines and two concentric ellipses. The ellipticity and PA of these
  two ellipses are those at $r=a_{\epsilon\ max}$. In the case of
  ``candidate bars'', the vertical lines have a dash-dot-dot-dot
  pattern. The horizontal dotted line and the white dotted ellipse
  mark the radius beyond which $\Delta \mu >
  0.2$\,mag\,arcsec$^{-2}$.}
\end{center}
\end{figure}

\newpage
\LongTables
\begin{deluxetable}{lrrrrrrrrr}
\setlength{\tabcolsep}{3pt}
\tabletypesize{\scriptsize}
\tablecolumns{10}
\tablecaption{Sample.\label{sample_tab}}
\tablewidth{0pt}
\tablehead{
\colhead{Galaxy} & \colhead{RA} & \colhead{Dec} & \colhead{Morph.} & \colhead{$T$} & \colhead{Dist.} & \colhead{$R_{25.5}$} & \colhead{$e_{25.5}$} & \colhead{$PA_{25.5}$} & \colhead{$M_{3.6\,\micron}$}\\
\colhead{} & \colhead{(hh:mm:ss)} & \colhead{(dd:mm:ss)} & \colhead{type} & \colhead{type} & \colhead{(Mpc)} & \colhead{($\arcsec$)} & \colhead{} & \colhead{(deg)} & \colhead{(AB mag)}\\
\colhead{(1)} & \colhead{(2)} & \colhead{(3)} & \colhead{(4)} & \colhead{(5)} & \colhead{(6)} & \colhead{(7)} & \colhead{(8)} & \colhead{(9)} & \colhead{(10)}}
\startdata
ESO026-001 & 20 24 59.0 & $-$81 34 35.5 &  SBc    &    5.9  &  19 &    57.3 & 0.060 & -61.9 & -18.01\\
ESO027-001 & 21 52 26.6 & $-$81 31 51.1 &  SBc    &    5.0  &  18 &   103.3 & 0.245 & -9.3 & -19.80\\
ESO079-007 & 00 50 04.2 & $-$66 33 07.8 &  Scd    &    7.0  &  25 &    54.9 & 0.182 & 19.7 & -18.43\\
ESO234-049 & 20 35 18.1 & $-$49 51 56.7 &  Sbc    &    4.0  &  36 &    59.4 & 0.130 & 58.6 & -19.70\\
ESO482-035 & 03 41 14.8 & $-$23 50 13.3 &  Sab    &    2.2  &  27 &    72.5 & 0.366 &  1.2 & -19.17\\
ESO549-018 & 03 48 14.1 & $-$21 28 27.3 &  SABc   &    4.9  &  26 &    87.8 & 0.385 & 15.0 & -19.44\\
IC0101 & 01 24 08.6 & 09 55 49.4 &  Sb     &    3.0  &  44 &    42.2 & 0.464 & -58.8 & -19.05\\
IC0167 & 01 51 08.6 & 21 54 45.9 &  SABc   &    5.0  &  28 &    76.3 & 0.463 & -66.8 & -18.64\\
\enddata
\tablecomments{Sample. (1): Galaxy name. (2), (3): RA and Dec of the nucleus, measured on the 3.6\,$\micron$ image. (4), (5): Morphological type and its numerical code, from LEDA. (6): Distance to the galaxy. We used the mean redshift-independent distance from NED when available, and the redshift-based one otherwise. (7), (8), (9): Radius along the semi-major axis, ellipticity ($1-b/a$) and position angle, measured at $\mu_{3.6\,\micron}=25.5$\,\maggie. (10): Asymptotic absolute magnitude at 3.6\,\micron.}
\end{deluxetable}

\newpage
\clearpage
\LongTables
\begin{landscape}
\begin{deluxetable}{lcrrrrrrrrrr}
\setlength{\tabcolsep}{3pt}
\tabletypesize{\scriptsize}
\tablecolumns{12}
\tablecaption{Profile measurements.\label{prof_measurements}}
\tablewidth{0pt}
\tablehead{
\colhead{Galaxy} & \colhead{Prof.\,type} & \colhead{$r_{1,i}$} & \colhead{$r_{2,i}$} & \colhead{$r_{1,o}$}& \colhead{$r_{2,o}$} & \colhead{$\mu_{0,i}$} & \colhead{$\mu_{0,o}$} & \colhead{$h_{i}$} & \colhead{$h_{o}$} & \colhead{$R_{\mathrm{br}}$} & \colhead{$\alpha$}\\
\colhead{} & \colhead{} & \colhead{(\arcsec)} & \colhead{(\arcsec)} & \colhead{(\arcsec)} & \colhead{(\arcsec)} & \colhead{(mag/\sq)} & \colhead{(mag/\sq)} & \colhead{(\arcsec)} & \colhead{(\arcsec)} & \colhead{(\arcsec)} & \colhead{(/\arcsec)}\\
\colhead{(1)} & \colhead{(2)} & \colhead{(3)} & \colhead{(4)} & \colhead{(5)} & \colhead{(6)} & \colhead{(7)} & \colhead{(8)} & \colhead{(9)} & \colhead{(10)} & \colhead{(11)} & \colhead{(12)}}
\startdata
ESO026-001 & 2i+2o &   31.27 &    46.82 &    50.31 &    60.62 & $   20.49^{+       0.09}_{-       0.05}$ & $   18.77^{+       0.34}_{-       0.56}$ & $   13.81^{+       0.45}_{-       0.28}$ & $    9.38^{+       0.69}_{-       0.90}$ & $   46.36^{+       0.65}_{-       0.42}$ & $    6.29^{+      11.18}_{-       2.03}$\\
ESO027-001 & 2o &   46.34 &    63.28 &    70.20 &    99.94 & $   20.82^{+       0.28}_{-       0.25}$ & $   17.22^{+       0.58}_{-       0.37}$ & $   32.68^{+       5.27}_{-       3.65}$ & $   13.41^{+       1.44}_{-       0.78}$ & $   75.20^{+       1.99}_{-       1.33}$ & $    0.72^{+       2.88}_{-       0.11}$\\
ESO079-007 & 1+(2o) &   17.37 &    57.30 & \nodata & \nodata & $   20.88^{+       0.06}_{-       0.10}$ & \nodata & $   13.33^{+       0.44}_{-       0.58}$ & \nodata & \nodata & \nodata\\
 ESO234-049 & 2o &   25.03 &    47.83 &    50.98 &    61.36 & $   20.54^{+       0.14}_{-       0.09}$ & $   18.37^{+       0.54}_{-       0.41}$ & $   14.41^{+       0.58}_{-       0.45}$ & $    8.99^{+       0.82}_{-       0.61}$ & $   47.88^{+       0.29}_{-       1.68}$ & $    5.08^{+      14.43}_{-       1.90}$\\
ESO482-035 & 2o &   27.19 &    55.48 &    59.02 &    72.28 & $   20.58^{+       0.03}_{-       0.05}$ & $   19.52^{+       0.77}_{-       0.40}$ & $   17.18^{+       0.30}_{-       0.33}$ & $   13.17^{+       2.63}_{-       1.05}$ & $   55.20^{+      21.18}_{-      13.40}$ & $    4.07^{+      10.05}_{-       1.50}$\\
ESO549-018 & 2o &   25.06 &    48.09 &    61.51 &    91.88 & $   21.43^{+       0.06}_{-       0.12}$ & $   18.63^{+       0.22}_{-       0.44}$ & $   44.88^{+       4.14}_{-       5.27}$ & $   13.76^{+       0.71}_{-       1.08}$ & $   51.22^{+       2.07}_{-       1.11}$ & $    0.37^{+       0.16}_{-       0.07}$\\
IC0101 & 1 &   14.86 &    45.65 & \nodata & \nodata & $   20.09^{+       0.09}_{-       0.10}$ & \nodata & $    8.43^{+       0.28}_{-       0.29}$ & \nodata & \nodata & \nodata\\
\enddata
\tablecomments{Structural properties of the disk profiles. (1): Galaxy name. (2): Profile type (see Sect.~\ref{classification}). Terms between brackets denote an uncertain classification. (3,4): Radial limits of the inner exponential. (5,6): Radial limits of the outer exponential. (7,8): Extrapolated central surface brightness of the inner and outer exponential. No inclination correction has been applied. (9,10): Radial scale-length of the inner and outer exponential. (11): Break radius. (12): Break sharpness: large values correspond to abrupt breaks and low values to smoother ones.}
\end{deluxetable}
\clearpage
\end{landscape}
\newpage
\clearpage
\LongTables
\begin{landscape}
\begin{deluxetable}{lrrrrrrrrrrrrrrrrrrrr}
\setlength{\tabcolsep}{3pt}
\tabletypesize{\scriptsize}
\tablecolumns{21}
\tablecaption{Bar properties.\label{bar_properties}}
\tablewidth{0pt}
\tablehead{
\colhead{} & \colhead{Bar} & \multicolumn{3}{c}{$\epsilon_{\mathrm{max}}$} & & \multicolumn{3}{c}{$\epsilon_{\mathrm{min}}$} & & \multicolumn{3}{c}{$\Delta \mathrm{PA} = 10^\circ$} & & \multicolumn{3}{c}{$\Delta \epsilon = 0.1$} & & \multicolumn{3}{c}{Final deprojected}\\
\cline{3-5} \cline{7-9} \cline{11-13} \cline{15-17} \cline{19-21}
\colhead{Galaxy} & \colhead{flag} & \colhead{$a(\arcsec)$} & \colhead{$\epsilon$} & \colhead{PA($^\circ$)} & & \colhead{$a(\arcsec)$} & \colhead{$\epsilon$} & \colhead{PA($^\circ$)} & & \colhead{$a(\arcsec)$} & \colhead{$\epsilon$} & \colhead{PA($^\circ$)} & & \colhead{$a(\arcsec)$} & \colhead{$\epsilon$} & \colhead{PA($^\circ$)} & & \colhead{$a(\arcsec)$} & \colhead{$\epsilon$} & \colhead{PA($^\circ$)}}
\startdata
ESO026-001 & C &     6 & 0.534 &  66.3 & &    16 & 0.478 &  61.7 & &   20.2 & 0.596 &  56.3 & &   26.4 & 0.434 &  22.1 & & $  11.5 \pm    5.2$ & 0.542 &  35.6\\
ESO027-001 & B &    18 & 0.533 &  59.4 & &    28 & 0.285 & 102.6 & &   23.2 & 0.464 &  69.4 & &   24.4 & 0.433 &  74.3 & & $  26.9 \pm    3.7$ & 0.628 & 165.4\\
ESO079-007 & B &    10 & 0.628 &  24.3 & &    16 & 0.150 &   5.6 & &   13.7 & 0.350 &  14.3 & &   12.4 & 0.528 &  24.7 & & $  11.2 \pm    1.2$ & 0.547 &   6.1\\
ESO482-035 & B &    20 & 0.723 &  28.5 & &    26 & 0.444 &  18.0 & &   25.8 & 0.447 &  18.5 & &   22.6 & 0.623 &  27.1 & & $  24.6 \pm    1.5$ & 0.673 &  41.9\\
ESO549-018 & C &    20 & 0.656 &  16.5 & &    26 & 0.169 &  39.2 & &   24.7 & 0.172 &  26.5 & &   22.3 & 0.556 &  18.6 & & $  21.2 \pm    1.2$ & 0.442 &   3.2\\
IC0167 & B &     6 & 0.487 & 137.5 & &     8 & 0.478 & 130.7 & &    9.1 & 0.503 & 127.5 & &   16.6 & 0.387 &  93.6 & & $   8.7 \pm    1.0$ & 0.430 & 147.3\\
IC0749 & C &    10 & 0.586 &  40.8 & &    16 & 0.139 &  23.5 & &   15.2 & 0.273 &  30.8 & &   13.6 & 0.486 &  41.7 & & $  13.2 \pm    2.0$ & 0.630 &   2.8\\
IC0800 & B &    20 & 0.779 & 149.0 & &    38 & 0.269 & 155.3 & &   55.0 & 0.291 & 159.0 & &   23.9 & 0.679 & 155.0 & & $  22.1 \pm    1.9$ & 0.721 & 165.5\\
\enddata
\tablecomments{Bar properties. The bar flag indicates whether we consider the elongated feature to be a bar (B) or just a candidate bar (C). For each bar we provide four measurements of its radius ($a$), its ellipticity ($\epsilon$) and its position angle (PA). The first set of measurements corresponds to the maximum ellipticity of the bar. The second one is determined at the local minimum in ellipticity after $\epsilon_{\mathrm{max}}$. The third group is measured at the point where the position angle departs by 10$^\circ$ from the one at $\epsilon_{\mathrm{max}}$. The fourth group of measurements correspond to the radius where $\epsilon$ drops by 0.1 with respect to $\epsilon_{\mathrm{max}}$. Note that the first two measurements of $a$ are integer multiples of 2\arcsec, as they are measured directly on the profiles at that resolution; the other two are interpolated at the corresponding $\Delta \epsilon$ and $\Delta \mathrm{PA}$. The last three columns show the finally adopted bar parameters, corrected for inclination.}
\end{deluxetable}
\clearpage
\end{landscape}

\end{document}